\documentclass[%
prfluids,
 reprint,
 onecolumn,
 superscriptaddress,
 aps,
 nofootinbib
]{revtex4-2}

\usepackage{silence}
\usepackage[T1]{fontenc}
\usepackage{graphicx}
\usepackage{dcolumn}
\usepackage{bm}
\usepackage{float}
\usepackage{amsmath,amssymb,amsfonts}
\usepackage{todonotes}
\usepackage{epstopdf}
\usepackage{multirow}

\def\XXint#1#2#3{{\setbox0=\hbox{$#1{#2#3}{\int}$}
\vcenter{\hbox{$#2#3$}}\kern-.5\wd0}}

\def\XXiint#1#2#3{{\setbox0=\hbox{$#1{#2{\mathrm{#3\!\! #3}}}{\iint}$}
\vcenter{\hbox{$#2 {\mathrm{#3\!\! #3}}$}}\kern-.5\wd0}}
\usepackage{enumitem}
\usepackage{hyperref}
\usepackage{color}  
\definecolor{darkGreen}{rgb}{0,0.45,0}
\definecolor{darkBlue}{rgb}{0,0,0.7}
\definecolor{darkRed}{rgb}{0.76, 0.13, 0.28}
\hypersetup{
	colorlinks=true, 
	linktoc=all,    
	linkcolor=darkBlue, 
	citecolor=darkGreen,
	urlcolor = darkRed,
}
\usepackage{comment}

\usepackage{overpic}

\renewcommand{\d}{{\mathrm{d}}}

\usepackage[margin=2.3cm]{geometry}
\begin{document}

\title{Modeling flying formations as flow-mediated matter}

\author{Christiana Mavroyiakoumou}%
\email[Electronic address: ]{cm4291@cims.nyu.edu}
\affiliation{Courant Institute, Applied Math Lab, New York University, New York, NY 10012, USA}

\author{Jiajie Wu}%
\affiliation{Courant Institute, Applied Math Lab, New York University, New York, NY 10012, USA}

\author{Leif Ristroph}%
\email[Electronic address: ]{ristroph@cims.nyu.edu}
\affiliation{Courant Institute, Applied Math Lab, New York University, New York, NY 10012, USA}

\date{\today}

\begin{abstract}
Collective locomotion of swimming and flying animals is fascinating in terms of individual-level fluid mechanics and group-level structure and dynamics. Here we bridge and relate these scales through a model of formation flight that views the collective as a material whose properties arise from the flow-mediated interactions among its members. We build on and revise an aerodynamic model describing how flapping flyers produce vortex wakes and how they are forced by others’ wakes. While simplistic, the model faithfully reproduces a series of physical experiments carried out over the last decade on pairwise interactions of flapping foils. By studying longer in-line arrays, we show that the group behaves as a soft  ``crystal’’ with regularly spaced member ``atoms’' whose positioning is, however, susceptible to deformations and dynamical instabilities. Poking or wiggling a member excites longitudinal waves (flow-mediated phonons, or ``flonons’’) that pass down the group while growing in amplitude, and indeed the internal excitation from flapping is sufficient to trigger instabilities. Linear analysis of the model explains the aerodynamic origin of the lattice spacing, the springiness of the ``bonds'' between flyers, and the tendency for disturbances to resonantly amplify. Other properties such as the timescales for instability growth and wave propagation seem to involve the full nonlinear behavior. These findings suggest intriguing analogies with physical materials that could be generally useful for understanding and analyzing animal groups. Several properties displayed by our system seem particularly relevant to biological collectives, namely group cohesion and organization, sensitive detection of and response to perturbations, and transmission of information through traveling waves.
\end{abstract}

\maketitle

\section{Introduction}

There are presently two schools of thought regarding how physics can be used to understand the collective movements of flying or swimming animals \cite{sumpter2006principles,cavagna2014bird,hemelrijk2012schools}. One class of models emphasizes how complex behaviors of the group can arise from simple behavioral rules of interaction among individuals. The classic example is the celebrated Vicsek model in which members moving at constant speed tend to align with neighbors \cite{vicsek1995novel}.
This work revealed a new type of phase transition in which disordered swarming gives way to directional flocking as orientational noise is decreased relative to the alignment effect. Much subsequent and ongoing research on related systems have sought to investigate collective dynamical properties that are not well described by conventional equilibrium statistical mechanics~\cite{vicsek2012collective,mora2016local}. The interaction rules are typically assumed rather than derived from other knowledge or measurements, and they may be interpreted as simplified representations of social behavior without account for physical interactions through flows. The second class of models emphasizes the fluid-dynamical interactions among members and how the group may collectively benefit through mechanisms such as energy savings, lift enhancement, or drag reduction~\cite{higdon1978induced,weihs1973hydromechanics,weihs1975some,lissaman1970formation,hummel1983aerodynamic}. Classic examples include the Lissaman-Schollenberger model of V-formation gliding flight of birds and Weihs' model of planar fish schools \cite{lissaman1970formation,weihs1973hydromechanics,weihs1975some}. Both involve interactions through vortex wakes, the former involving tip vortices trailing the wings and the latter arrays of eddies shed by oscillating tail fins. These and many subsequent studies consider idealized systems such as orderly lattices of individuals fixed at relative positions and moving synchronously, with the goal being to determine fluid-mechanically optimal arrangements \cite{liao2007review,partridge1979evidence,pavlov2000patterns,badgerow1981energy,abrahams1987fish,larsson2012fish}.

Can these two views be reconciled? A bridge may be provided by Sir James Lighthill's proposal from the 1970s that the structures of animal collectives might arise in part or perhaps dominantly from the flow-mediated forces of interaction among the group members \cite{lighthill1975mathematical}. Uncoordinated swarming or milling of fish at low swimming speeds may give way to orderly lattices of strongly aligned members who are forced into position by the strong vortex wakes generated at high speeds. This picture is analogous to atoms or molecules that transition from a disordered gas phase to an orderly crystalline solid as the interactions are strengthened relative to thermal agitation. The so-called Lighthill conjecture has been revived over the last decade \cite{ramananarivo2016flow}, during which it has inspired many investigations into the fluidic forces that are in any case present in schools and flocks and should be better characterized if these phenomena are to be understood~\cite{zhu2014flow,becker2015hydrodynamic,ramananarivo2016flow,dai2018stable,peng2018hydrodynamic,newbolt2019flow,newbolt2024flow}. We see Lighthill's picture as harmonizing ideas emphasized by the two schools: Local interactions lead to emergent group structures and motions, and fluid mechanical interactions specifically may provide the benefit of organizing the collective state.

Our goals in this work are to formulate a dynamical model of collective locomotion based on fluid mechanical interactions and to use its predictions to explore Lighthill's view of the group as a material bound together by physical forces. We focus on in-line, columnar, series or tandem formations in which flyers are arranged one after another and whose motions along this line and relative positions within it are dynamically determined by the flow interactions. This idealized context has been recognized in recent studies as being the simplest setting for studying wake interactions among self-propelling bodies and their effects on the group structure and dynamics~\cite{hemelrijk2012schools,becker2015hydrodynamic,peng2018hydrodynamic,oza2019lattices,saadat2021hydrodynamic,heydari2021school,newbolt2024flow,nitsche2024stability}. We further idealize the interactions as dominantly occurring through the flows generated by a vertically oscillating wing or foil, which is considered emblematic of high Reynolds number flapping locomotion~\cite{triantafyllou1993optimal,andersen2017wake,becker2015hydrodynamic,ramananarivo2016flow}.

The problem of flapping flight in linear formations also allows for detailed validation against laboratory experiments and fluid dynamical simulations conducted over the last decade on actuated foils, wings or filaments. Some such efforts take the form of so-called robophysical experiments that involve mechanically actuated hydrofoils or airfoils as analogues for fins or wings~\cite{becker2015hydrodynamic,ramananarivo2016flow,newbolt2019flow}.
Previous works considered the thrust generation and efficiency of a single flapping foil~\cite{triantafyllou1993optimal} or a pair of flapping foils in tandem~\cite{boschitsch2014propulsive,rival2011recovery,han2025tailoring} fixed within an oncoming flow.
Other studies have focused on the free locomotion of flapping foils in forward flight~\cite{vandenberghe2004symmetry,alben2005coherent,vandenberghe2006unidirectional,spagnolie2010surprising,ramananarivo2016flow,newbolt2019flow,newbolt2024flow}. In these systems, forward locomotion emerges as a consequence of the interaction with the surrounding fluid, and hence the collective dynamics of multiple such flyers emerges from their flow interactions. 
Because it allows precise control and measurement of motions, forces, and flows, this line of work has advanced our understanding of the mechanisms underlying flapping locomotion for foils operating at biologically relevant conditions. The approach emphasizing emergent dynamics also shares some connections with active matter physics, which considers how energy-consuming constituents interact to organize the collective state of a system. The limit of large system sizes, however, seems beyond the reach of computational fluid dynamical simulations and lab experiments, and hence modeling can play a central role.

Here we propose a mathematical model intended to capture the general features of collective and interactive locomotion dynamics and which can scale up to arbitrarily large groups. We build on and revise an aerodynamic model that describes the fluid-structure interactions among the flyers through a simplified representation of the wake flow generated by their flapping motions~\cite{newbolt2019flow,newbolt2024flow}. Our work below is organized according to the formulation of the model, validation against previous experiments, analysis relevant to small systems, and numerical investigations into larger systems. The key results are that the model replicates previous results with high fidelity and further shows that collectively locomoting systems have strong connections with states of matter and material systems, including crystalline structures, elastic response and longitudinal waves, and plastic deformation and failure.

\section{Follower-wake interaction model}\label{sec:model}

We begin by formulating a dynamical model of collective locomotion based on fluid mechanical interactions. Two system geometries, or equivalently simulation boundary conditions, are considered: One representing the group flying in a semi-infinite quiescent fluid and another modeling flight in a periodic domain, which resembles experimental conditions. To explore the general characteristics of long chains of flapping flyers in high-Reynolds-number flows across a wide range of the relevant parameters, we also non-dimensionalize the governing equations. Finally, to reveal the structural and dynamical properties of flying formations, we consider the effects of various types of external forcing.

\subsection{Formulation of the dynamical model}
\begin{figure}[htbp!]
    \centering
    \hspace*{-.75cm}
    \includegraphics[width=.82\textwidth]{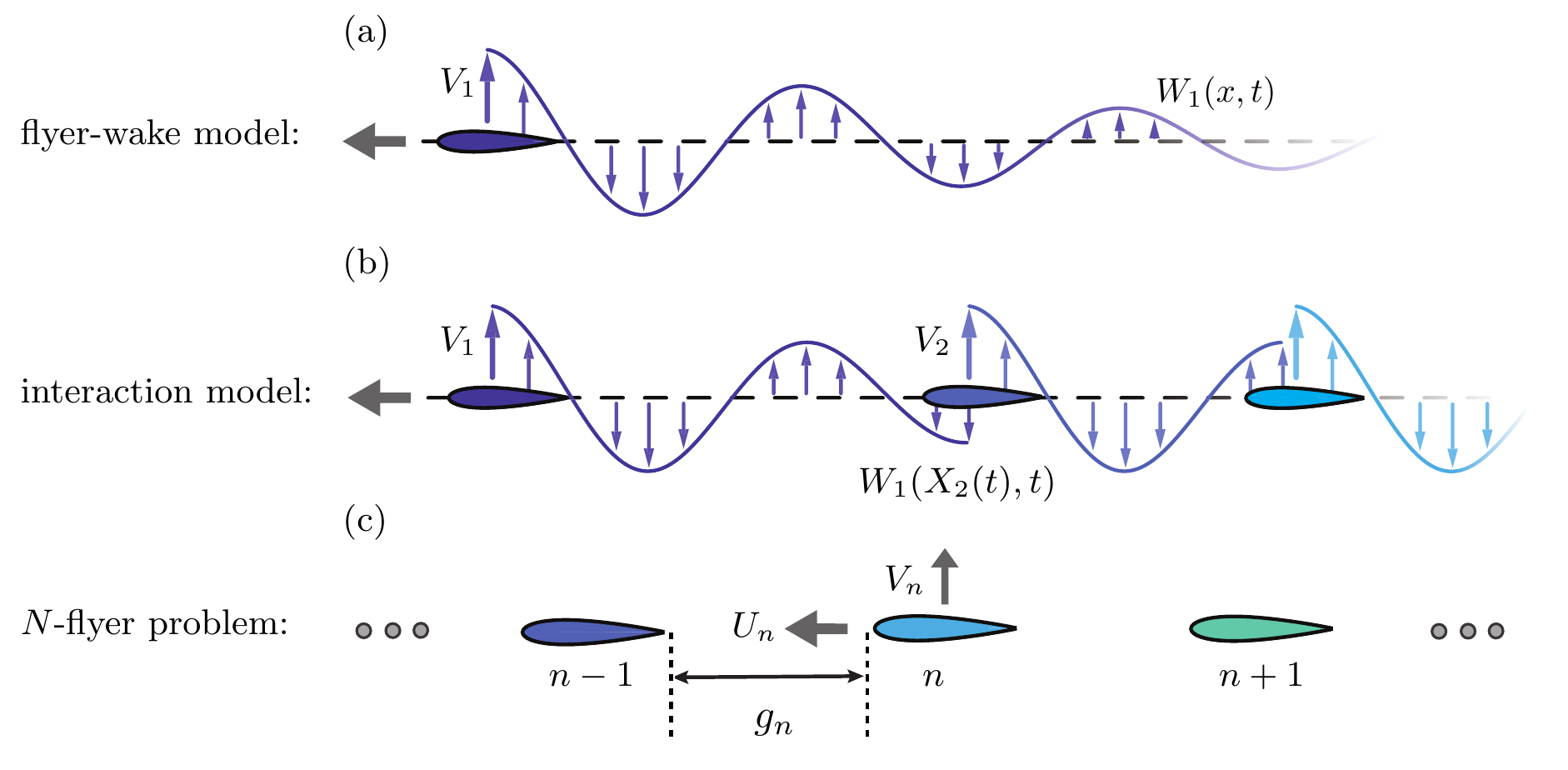}
    \caption{
    Schematic diagrams of a model of wake generation and interaction. (a) A flyer emits a wake whose speed directly reflects its flapping speed $V_1$, and the flow thereafter decays in time. (b) A model to simulate the group dynamics with one-way, nonreciprocal interactions that have memory. The next flyer experiences thrust that depends on its flapping speed $V_2$ relative to the ambient wake $W_1(X_2(t),t)$. (c) Idealized problem of a linear flying formation of $N$ flapping flyers indexed by $n$, each with prescribed vertical oscillations $V_n$ and free forward flight motions $U_n$. Here $g_n=X_{n-1}-X_n$ is the gap distance between flyer $n$ and its upstream neighbor $n-1$.}
    \label{fig:modelSchem}
\end{figure}

We build on and revise related dynamical models, in which each flyer moves along a line while emitting a wake flow and interacting with the wake flows emitted by others~\cite{becker2015hydrodynamic,newbolt2019flow,newbolt2024flow}. In particular, each flyer has  prescribed flapping and this dictates the flyer’s self-propulsion as well as the wake flow signal left behind in its trail, as shown schematically in Fig.~\ref{fig:modelSchem}(a). 
The flyer experiences a propulsive force that depends on how its instantaneous flapping signal interferes with the ambient wake signal left by others (see Fig.~\ref{fig:modelSchem}(b)). This general framework addresses the fact that flocking involves interactions through long-lived flows, which retain memory of the conditions under which they were generated~\cite{becker2015hydrodynamic}.

We treat each flyer as an inertial body free to move in the horizontal direction due to aerodynamic interactions. 
The flyers have instantaneous, horizontal positions~$X_n(t)$, flight speeds $U_n(t)=\dot{X}_n(t)$, and prescribed vertical oscillations (heaving-and-plunging motions) with speed $V_n(t)=\pi A_n f_n\cos(2\pi f_n t)$, where $f_n$ is the flapping frequency and~$A_n$ is the peak-to-peak flapping amplitude. Here, $n=1,2,\dots, N$ is used to label each of the $N$ flyers in the group, with $n=1$ representing the leader.

The horizontal dynamics of each individual are governed by Newton's second law of motion:
\begin{equation}\label{eq:N2L}
    M_n\dot{U}_n(t)=T_n(t)-D_n(t),\quad n=1,2,\dots, N,
\end{equation}
where $M_n$ is the mass of flyer $n$, $T_n(t)$ is the horizontal thrust force, and $D_n(t)$ is a skin friction drag. For simplicity, we assume that all flyers have the same mass, $M_n=M$.
In the absence of any interactions, these aerodynamic forces take the forms $\rho C_TcsV_n^2/2$ and $C_Ds\sqrt{\rho\mu c}\, U_n^{3/2}/2$, respectively, with thrust varying quadratically with the flapping (vertical) speed~\cite{floryan2017scaling,smits2019undulatory,heydari2021school} and drag as the $3/2$-power of the propulsion (horizontal) speed. 
Here $\rho$ is the fluid density, $\mu$ is its dynamic viscosity, $c$ and $s$ are the wing's chord and span length, and $C_T$ and $C_D$ are the dimensionless coefficients of thrust and drag, respectively~\cite{tritton2012physical,vandenberghe2006unidirectional,vandenberghe2004symmetry}.
The skin friction drag along the upper and lower flyer surfaces is modeled using Blasius laminar boundary layer theory~\cite{schlichting2016boundary} and is derived through $\rho  (C_D/\sqrt{\mathrm{Re}})cs U_n^{2}/2$ with the Reynolds number $\mathrm{Re}=\rho U_nc/\mu$. 
It is crucial that the drag force follows a $3/2$-power law on the flyer's  horizontal speed, instead of a quadratic power law used in previous works~\cite{newbolt2019flow,newbolt2024flow}. This ensures that hysteresis loops observed experimentally, showing multiple stable modes for the same kinematic parameters~\cite{becker2015hydrodynamic}, are recovered.

The aerodynamic interactions are captured by a relative flow
model in which the thrust is modified to depend on the vertical speed of the wing relative to any ambient wake signal, $\Delta V_{n}(t)$. Specifically, flyer $n$ experiences a thrust proportional to the square of $\Delta V_{n}(t)=V_{n}(t)-W_{n-1}(X_{n}(t),t)$. The flow speed of the wake emitted by flyer $n$ is denoted by $W_n(x,t)$ and is a function of space and time, whose value at the flyer's location is assumed to be equal to the flapping speed ${W_{n}(X_n(t),t)=V_{n}(t)}$~\cite{becker2015hydrodynamic,ramananarivo2016flow,newbolt2019flow,oza2019lattices,newbolt2024flow}.
Here we assume an ``erase-and-replace'' scheme in which each flyer overwrites the wake signal of its upstream neighbor with its own signal, as shown in Fig.~\ref{fig:modelSchem}(b). Therefore, the interactions are nearest-neighbor and one-way in the downstream direction: Each follower in a pair is forced by the leader, but cannot force it~\cite{newbolt2024flow}. 

The flyer located at $X_{n-1}(t_n(t))$ generates a signal in the form of a wake with flow velocity given by
$V_{n-1}(t_n(t))e^{-(t-t_n(t))/\tau}$. These terms involve the earlier time $t_n(t)$ when the upstream neighbor $n-1$ was at the current position of flyer $n$, as defined through the implicit relation $X_n(t)=X_{n-1}(t_n(t))$. Thus, $t_n(t)$ describes the effect of memory in the system and $t-t_n(t)$ denotes the delay time between when a signal left by flyer $n-1$ is encountered by flyer $n$. This signal decays exponentially in time with timescale~$\tau$, corresponding to wake dissipation~\cite{ramananarivo2016flow}. For the model validation with past robophysical experiments with flapping foils in water~\cite{becker2015hydrodynamic,ramananarivo2016flow,newbolt2019flow,newbolt2024flow}, discussed later in Sec.~\ref{sec:validationWithExperiments}, we use $\tau=0.5$ s, which corresponds to the value specifically inferred in these experimental systems. However, to examine larger groups in Sec.~\ref{sec:manybodies},  we use $\tau\gg 1$, corresponding to a much slower dissipation rate and thus stronger interactions.
The memory timescale $t_n$ is itself time-dependent, motivating its inclusion as a state variable in the dynamical system, by deriving its evolution equation by the chain rule: $\dot{X}_n(t)=\dot{X}_{n-1}(t_n(t))\dot{t}_n(t)$. This results in ${\dot{t}_n(t)=\dot{X}_n(t)/\dot{X}_{n-1}(t_n(t))=U_n(t)/U_{n-1}(t_n(t))}$, where the term $U_{n-1}(t_n(t))$ constitutes a state-dependent delay.

The propulsion dynamics for each flyer are thus described by a system of nonlinear delay differential equations for $(X_n, U_n,t_n)$, given by:
\begin{align}
    \dot{X}_n(t)&=U_n(t),\label{eq:xdot}\\
    \dot{U}_n(t)&=\frac{\rho C_Tcs}{2M}\left[V_n(t)-V_{n-1}(t_n(t))e^{-(t-t_n(t))/\tau}\right]^2-\frac{C_Ds\sqrt{\rho \mu c}}{2M} U_n^{3/2}(t),\label{eq:UndotDimensional}\\
    \dot{t}_n(t)&=\frac{U_n(t)}{U_{n-1}(t_n(t))}.\label{eq:tnsystem}
\end{align}
A group of $N$ flyers is governed by $3N$ such equations. Simulations are performed using \textsc{Matlab}'s \texttt{ddesd} solver for state-dependent delay differential equations~\cite{MATLAB}.

\subsection{System geometry and boundary conditions}
In this work, we consider two configurations: open and closed. The open configuration corresponds to the ensemble flying in a semi-infinite domain of quiescent fluid, while the closed setup represents flight within a cyclic domain, as in the robophysical experiments. The treatment of the leader $n=1$, varies depending on the chosen configuration. If open, the leader swims exactly as a solo flyer in a linear formation. This means that the interaction term involving $V_{n-1}$ in Eq.~\eqref{eq:UndotDimensional} is removed, as is Eq.~\eqref{eq:tnsystem} entirely. If closed, the leader interacts with the last member in the group since each flyer swims within the wake of the other. For instance, for a two-flyer system moving in a closed, cyclic geometry, Eq.~\eqref{eq:UndotDimensional} becomes:
\begin{equation}\label{eq:ramananarivoU}
    \dot{U}_{1,2}(t)=\frac{\rho C_Tcs}{2M}\left[V_{1,2}(t)-V_{2,1}(t_{1,2}(t))e^{-(t-t_{1,2}(t))/\tau}\right]^2-\frac{C_Ds\sqrt{\rho \mu c}}{2M} U_{1,2}^{3/2}(t), 
\end{equation}
where the subscript 1 refers to the leader and the subscript 2 refers to the follower.
In the closed configuration, the results depend on the length $C$ of the cyclic domain, which determines the spacing between the individuals. In particular, in Eq.~\eqref{eq:ramananarivoU} the parameter $C$ appears implicitly within the memory- or history-dependent thrust term, and it dictates the gap between the follower and the leader. This gap is equal to ${C-g=C-(X_1-X_2)}$. In general, smaller values of $C$ result in stronger interactions between the flyers, while $C\to \infty$ resembles an open configuration, where the wake flow of the last member has a negligible effect on the leader. As the number of flyers increases, the spacing between them in the array decreases, leading to amplified flow interactions.

If the setup involves a single flyer flying in a closed, cyclic configuration, interacting with its own wake, the system of delay differential equations, given in Eqs.~\eqref{eq:xdot}--\eqref{eq:tnsystem}, becomes
\begin{equation}\label{eq:beckerUndot}
    \dot{X}(t)=U(t)\quad ;\quad
    \dot{U}(t)=\frac{\rho C_Tcs}{2M}\left[V(t)-V(t_0(t))e^{-(t-t_0(t))/\tau}\right]^2-\frac{C_Ds\sqrt{\rho \mu c}}{2M} U^{3/2}(t)\quad;\quad
    \dot{t}_0(t)=\frac{U(t)}{U(t_0(t))}.
\end{equation}

As in previous studies~\cite{becker2015hydrodynamic,ramananarivo2016flow,newbolt2019flow}, to characterize arrays of flyers we define the dimensionless spacing ${S_n=g_n/\lambda_{n-1}}$, where $g_n=X_{n-1}-X_n$ is the gap distance between flyer $n$ and its upstream neighbor $n-1$ (illustrated schematically in Fig.~\ref{fig:modelSchem}(c)), and $\lambda_{n-1}=U_{n-1}/f_{n-1}$ is the wavelength of the undulatory trajectory of the leading member in each pair of consecutive flyers, formed by the flapping motion. 

\subsection{Model non-dimensionalization}\label{sec:modelDimensionless}

To characterize the collective dynamics of long chains of flapping flyers, we recast our model in dimensionless form and consider $\tau \to \infty$ for the wake decay time constant. This limit corresponds to long-lived inertial flows, which is the regime most relevant to the interactions involved in high-$\mathrm{Re}$ flocking behavior. 

There are five relevant physical and kinematic quantities associated with each flyer in a group: chord length $c$ [cm], span $s$ [cm], mass $M$ [g], flapping amplitude $A$ [cm] and flapping frequency $f$ [1/s], and two physical quantities relevant to the fluid: density $\rho$ [g cm$^{-3}$] and dynamic viscosity~$\mu$ [g cm$^{-1}$s$^{-1}$].
We non-dimensionalize the governing equations (Eqs.~\eqref{eq:xdot}--\eqref{eq:tnsystem}) by the flapping amplitude $A$, the characteristic timescale~$1/f$, and the typical flapping speed $Af$. In particular, we use 
\begin{equation}\label{eq:tildeQuantities}
    \tilde{t}= ft,\quad  \tilde{X}_n=\frac{X_n}{A},\quad \tilde{U}_n=\frac{U_n}{Af}.
\end{equation}
The dimensionless forms of Eqs.~\eqref{eq:xdot} and~\eqref{eq:tnsystem}
are the same as before, dropping tildes. With the $\tau\to \infty$ assumption, the term representing the exponential decay of flow speed in Eq.~\eqref{eq:UndotDimensional} is dropped, and Eq.~\eqref{eq:UndotDimensional} becomes
\begin{equation}\label{eq:nondimensionalizingUndot}
    Af^2\dfrac{\d \tilde{U}_n}{\d \tilde{t}}=\frac{\rho C_T cs(\pi  A f)^2}{2M}\left[\cos(2\pi\tilde{t})-\cos(2\pi\tilde{t}_n(\tilde{t}))\right]^2-\frac{C_Ds\sqrt{\rho \mu c}}{2M}(Af)^{3/2}\tilde{U}_n^{3/2},
\end{equation}
with dimensionless quantities (and their dimensionless derivatives) denoted by tildes.
Dividing Eq.~\eqref{eq:nondimensionalizingUndot} by $Af^2$, the dimensionless equations are
\begin{align}
    \dot{\tilde{X}}_n(\tilde{t})&=\tilde{U}_n(\tilde{t}),\label{eq:XndotDimensionless}\\
    \dot{\tilde{U}}_n(\tilde{t})&=\frac{C_T\pi^2 A^*}{2M^*}\left[\cos(2\pi \tilde{t})-\cos(2\pi \tilde{t}_n(\tilde{t}))\right]^2-\frac{C_DA^*}{2M^*}\frac{1}{\sqrt{\mathrm{Re}_f}} \tilde{U}_n^{3/2}(\tilde{t}),\label{eq:UndotDimensionless}\\
    \dot{\tilde{t}}_n(\tilde{t})&=\frac{\tilde{U}_n(\tilde{t})}{\tilde{U}_{n-1}(\tilde{t}_n(\tilde{t}))}.\label{eq:TndotDimensionless}
\end{align}
The dimensionless groups are:
\begin{equation}\label{eq:4dimGroups}
   M^*=\frac{M}{\rho c^2s},\quad \mathrm{Re}_f=\frac{\rho A fc}{\mu},\quad A^*=\frac{A}{c}, \quad s^*=\frac{s}{c},\quad C_T, \quad C_D,
\end{equation}
where $M^*$ is the dimensionless mass, $\mathrm{Re}_f$ is the flapping Reynolds number, $A^*$ is the flapping oscillation amplitude, $s^*$ is the planform aspect ratio, $C_T$ is the coefficient of thrust, and $C_D$ is the coefficient of skin friction drag. We note that $s^*$ does not appear explicitly in our system and the coefficients  $C_T$ and $C_D$ are set as 1 and~10, respectively. 
Previous measurements~\cite{triantafyllou1993optimal,floryan2017scaling} indicate that $C_T\approx 0.8$ to 1.1 after non-dimensionalization in terms of the flapping speed. Here, we select $C_T=1$ for simplicity and to be consistent with experiments~\cite{newbolt2019flow,newbolt2024flow}. Details regarding the choice of $C_D=10$ can be found in Sec.~\ref{sec:solo}.

\subsection{Types of external forcing}\label{sec:forcePerturbations}

Applying forces will prove useful for revealing the structural and dynamical properties of formations.
In this work, we consider three types of external forces to be applied to individuals:
\begin{align}
   F_{\mathrm{DC}}(t) &= F,&&(\text{steady/DC force})\label{eq:dcperturbation}\\
    F_{\mathrm{AC}}(t) &= a_{\mathrm{AC}}\sin(2\pi f_{\mathrm{AC}}t-\phi),&& (\text{oscillatory/AC force})\label{eq:acperturbation}\\
    F_I(t)&=-\frac{F_1}{\sqrt{2\pi\sigma^2}}\frac{e^{-(t-t_0)^2/(2\sigma^2)}}{\Phi(\delta/(2\sigma))-\Phi(-\delta/(2\sigma))}.&& (\text{``kick''/impulse force})\label{eq:ImpulseForceGaussian}
\end{align}

The first type of external forcing is a steady or DC force (Eq.~\eqref{eq:dcperturbation}). To incorporate this in the model, a constant term $F/M$ is added to the right-hand side of the $\dot{U}_n$ equation (Eq.~\eqref{eq:UndotDimensional}). Here, we adopt the same convention for the force direction as previous works~\cite{ramananarivo2016flow,newbolt2024flow}: Positive values of $F$ ($F>0$) drive the flyer toward its upstream neighbor, while negative values ($F<0$) push the flyer away from its upstream neighbor. 

The second kind of external forcing is an oscillatory or AC perturbation (Eq.~\eqref{eq:acperturbation}). This is accomplished by adding a term of the form $F_{\mathrm{AC}}(t)/M$ to the right-hand side of the $\dot{U}_n$ equation. In Eq.~\eqref{eq:acperturbation}, $a_{\mathrm{AC}}$ and $f_{\mathrm{AC}}$ represent the amplitude and frequency of the perturbation, respectively, and $\phi$ denotes the phase lag between the flyer's flapping motion and the applied perturbation. 

Lastly, an impulse force in the form of a truncated Gaussian bump, described by Eq.~\eqref{eq:ImpulseForceGaussian}, is also considered.
This perturbation is not persistent in time. It corresponds to a ``kick'' at $t\approx t_0$ whose duration is set by the parameter $\delta$ and whose magnitude is determined by $F_1$ in Eq.~\eqref{eq:ImpulseForceGaussian}. Here, $t_0$ is the time at which the impulse force is at its maximum, $\sigma$ is the 
standard deviation of a normal distribution without truncation, and $\Phi$ is the cumulative distribution function of the standard normal distribution. The normalization by $\Phi(\delta/(2\sigma))-\Phi(-\delta/(2\sigma))$ ensures that the truncated Gaussian bump remains a valid distribution. As with the steady DC force, the minus sign indicates that the applied force drives the flyer toward its downstream neighbor.

\section{Model validation with robophysical experiments}\label{sec:validationWithExperiments}

Numerous robophysical experiments carried out over the last decade have aimed to explain the interplay between physical and behavioral interactions in formation locomotion~\cite{becker2015hydrodynamic,ramananarivo2016flow,newbolt2019flow,newbolt2024flow}. Here, to validate the follower-wake interaction model, we compare the results from these experiments with the simulation results obtained by solving the dimensional equations~\eqref{eq:xdot}--\eqref{eq:tnsystem}. By using both  numerics and theory, we also find new phenomena within these previous setups. This section is organized in order of more degrees of freedom in the system and increasing group sizes. 

\begin{figure}[htbp!]
 \includegraphics[width=.93\textwidth]{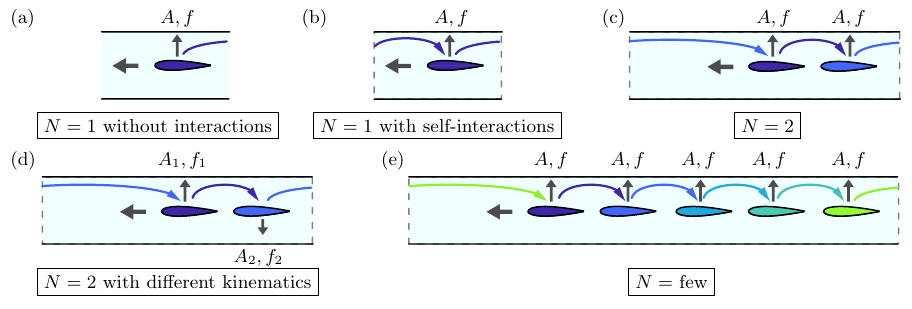}
    \caption{Schematic diagrams of the experimental setups used to validate the follower-wake interaction model in Secs.~\ref{sec:solo}--\ref{sec:newbolt2022}: (a)~${N=1}$ without interactions, (b) $N=1$  with self-interactions~\cite{becker2015hydrodynamic}, (c) $N=2$~\cite{ramananarivo2016flow}, (d) $N=2$ with different~kinematics~\cite{newbolt2019flow}, and (e) $N=$ few flyers~\cite{newbolt2024flow}. In the simulations, the closed configuration is represented by cyclic boundary conditions. With the exception of panel (d), the flyers in all other setups have identical flapping motions.}
    \label{fig:expValidationSchem}
\end{figure}

Four past experiments are used for comparison, each briefly summarized below, along with an investigation of the simplest case possible, which is a single isolated flyer flying in a semi-infinite domain of quiescent fluid in the absence of any interactions (see Fig.~\ref{fig:expValidationSchem}(a)). The second setup is a single-flyer problem, where the flyer is flying in circular orbits~\cite{becker2015hydrodynamic}, interacting with its own wake while undergoing repeated passes across a domain specified by cyclic boundary conditions, as shown schematically in Fig.~\ref{fig:expValidationSchem}(b).
In the simulations, the cyclic boundary conditions mimic the closed, circular geometry of the experimental setup. These are represented by dashed lines at both ends of the domain in Figs.~\ref{fig:expValidationSchem}(b)--(e), with ``flow'' arrows leaving the last member and reaching the leader.
In Fig.~\ref{fig:expValidationSchem}(b), the formation can be viewed as being forced to have a certain periodicity, similar to the Weihs-type models, where the  formation is fixed and the benefits of the group are determined in terms of energetics~\cite{weihs1973hydromechanics,weihs1975some,oza2019lattices,fish1991burst,hemelrijk2012schools}.
The remaining three setups, shown schematically in Figs.~\ref{fig:expValidationSchem}(c)--(e), make use of an important concept from the Vicsek-type models, where the formation is allowed to spontaneously emerge and the bodies self-organize either by attracting or repelling each other. The system illustrated in Fig.~\ref{fig:expValidationSchem}(c), involves $N=2$ flyers with identical flapping motions and with the gap spacing between them emerging as a result of their flow interactions~\cite{ramananarivo2016flow}. Figure~\ref{fig:expValidationSchem}(d) depicts two individuals with kinematic variability, meaning that they can flap synchronously or asynchronously and can have different flapping amplitudes and frequencies~\cite{newbolt2019flow}. Finally, Fig.~\ref{fig:expValidationSchem}(e) shows a schematic for a few-flyer system where the flyers have identical flapping motions~\cite{newbolt2024flow}. 

\subsection{Single isolated flyer problem}\label{sec:solo}

The dynamical model is first applied on the simplest possible setup, which is a single, non-interacting flyer (see schematic in Fig.~\ref{fig:expValidationSchem}(a)). Using the open configuration, we derive the skin friction drag coefficient, compare with past experimental measurements, and analyze the equilibrium flight dynamics. 

The skin friction drag on the flyer's upper and lower surfaces is modeled using Blasius laminar boundary layer theory~\cite{schlichting2016boundary,tietjens1957applied,white1966fluid,fang2025flowinteractionsforwardflight}. Therefore, the drag force in Eq.~\eqref{eq:N2L} is of the form $D=\rho C_D^*(\mathrm{Re})cs U^2/2$, where $C_D^*(\mathrm{Re})=C_D/\sqrt{\mathrm{Re}}$ is the coefficient of drag, $\mathrm{Re}=\rho Uc/\mu$ is the Reynolds number, and $C_D$ is a constant to be determined.

A single, non-interacting flyer moving at equilibrium experiences a thrust force that is balanced by a drag force:
\begin{equation}\label{eq:skin2.6}
    \frac{\rho C_Tcs}{2M}V^2-\frac{C_Ds \sqrt{\rho\mu c}}{2M}U^{3/2}=0.
\end{equation}
The vertical flapping speed $V$ of the isolated flyer is given by ${V=\pi Af \cos(2\pi f t)}$, and so the period-averaged vertical speed is $\langle V^2\rangle_{T_f} =(\pi Af)^2/2$. Upon rearrangement and using $\langle U\rangle_{T_f}=U^*$, the equilibrium flight speed for a single flyer can be written as:
\begin{equation}\label{eq:Ueqm}
    U^*=\left( \frac{\rho c}{\mu}\right)^{1/3} \left(\frac{C_T}{2C_D} \right)^{2/3} (\pi  Af)^{4/3}.
\end{equation}
Past experiments involving flapping foils in water~\cite{ramananarivo2016flow} serve as our benchmark and provide the following values for our model: fluid density $\rho=1\,\mathrm{g}/\mathrm{cm}^3$, fluid viscosity $\mu=0.01\,\mathrm{g}/(\mathrm{cm}\,\, \mathrm{s})$, and foil chord length $c=3.8$~cm. Substituting these in Eq.~\eqref{eq:Ueqm} results in $U^*\approx{21}(C_T/C_D)^{2/3}(Af)^{4/3}$, which yields the following scaling law: 
    $U\sim K(Af)^{4/3}$.
By fitting the data from the flight speed of an isolated flyer in experiments~\cite[Fig.~2(b)]{ramananarivo2016flow} we can fix the ratio $C_T/C_D$ in our model to account for flight speeds at all $A$ and $f$. The model predicts $C_D\approx 10$ with $C_T=1$ as the thrust coefficient~\cite{triantafyllou1993optimal,floryan2017scaling}.
Therefore, $C_D^*(\mathrm{Re})\approx 10/\sqrt{\mathrm{Re}}$ and the drag force  written explicitly in terms of the emergent flight speed is $D=5s\sqrt{\rho\mu c}\,U^{3/2}$.

\begin{figure}[htbp!]
    \centering
    \includegraphics[width=\linewidth]{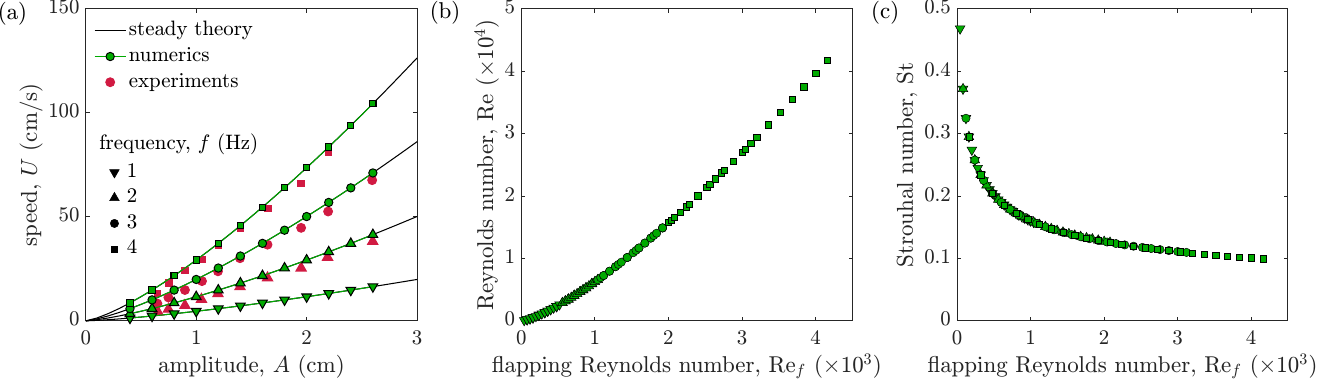}
    \caption{Propulsion dynamics of a single isolated flyer. (a) Emergent flight speed $U$ for motions of varying $A$ and $f$, compared with the theoretical equilibrium speed~$U^*$ from Eq.~\eqref{eq:Ueqm} (solid black lines) and experimental data (red markers) of the average propulsion speed of a flapping foil propelling in orbits around a cylindrical water tank~\cite[Fig.~2(b)]{ramananarivo2016flow}. (b)~Flight speed in dimensionless form given as Reynolds number versus flapping Reynolds number,  which captures the flapping kinematics. (c) Strouhal number $\mathrm{St}$, as defined in Eq.~\eqref{eq:strouhal}, plotted as a function of $\mathrm{Re}_f$.}    \label{fig:soloFlyerDynamics}
\end{figure}

With all the terms in the governing equations now fully determined, model predictions of emergent flight dynamics are made for the simplest case: a single, non-interacting flyer flying in an open configuration. Figure~\ref{fig:soloFlyerDynamics}(a) shows that the flight speed increases with both the flapping amplitude and frequency. Experimental data of the average propulsion speed of a single flapping foil propelling in orbits around a cylindrical water tank are overlaid for direct comparison as red markers for the corresponding kinematic parameters~\cite[Fig.~2(b)]{ramananarivo2016flow}. The agreement is excellent: the numerical flight speeds $U$ (green markers) closely match both the experimental data and the theoretical equilibrium speed $U^*$ given by Eq.~\eqref{eq:Ueqm} and shown as solid black lines. 
The nonlinear response observed in Fig.~\ref{fig:soloFlyerDynamics}(a) is our primary motivation for modifying the models used in prior studies, where the drag force was attributed to pressure drag scaling with $U^2$~\cite{newbolt2019flow,newbolt2024flow}. In contrast, the current work involves skin friction drag, which scales as $U^{3/2}$. Balancing thrust and drag forces in this case yields $(Af)^2\sim U^{3/2}$, leading to the nonlinear scaling relationship $U\sim (Af)^{4/3}$.

The dependence of the Reynolds number $\mathrm{Re}=\rho Uc/\mu$, which characterizes flight speed, on the flapping Reynolds number $\mathrm{Re}_f=\rho A f c/\mu$, a parameter that captures the effects of flapping kinematics, is shown in Fig.~\ref{fig:soloFlyerDynamics}(b).
The monotonically increasing trend highlights how variations in the flapping parameters influence the resulting aerodynamics. A key dimensionless parameter known as the Strouhal number, defined by
\begin{equation}\label{eq:strouhal}
    \mathrm{St}=\frac{Af}{U} = \frac{\mathrm{Re}_f}{\mathrm{Re}}, 
\end{equation}
is used to characterize propulsive performance in flows associated with bird-like flight and fish-like swimming, as it compares the input effort of the individual characterized by the flapping speed $Af$ to the output represented by the propulsion speed $U$~\cite{triantafyllou1993optimal,eloy2012optimal,sfakiotakis1999review,taylor2003flying}. 
In Fig.~\ref{fig:soloFlyerDynamics}(c), we show the Strouhal number as a function of $\mathrm{Re}_f$. The observed values are between 0.1 and 0.5 and span the optimal range $\mathrm{St} \in (0.2, 0.4)$ identified in high-Re studies of oscillating wings~\cite{triantafyllou1993optimal}.

\subsection{Self-interacting flyer as an infinite lattice of fixed spacing}\label{sec:becker}

The model is now validated using the single, self-interacting flyer problem, shown in Fig.~\ref{fig:expValidationSchem}(b). 
In this setup, the flyer undergoes prescribed flapping motions with frequency~$f$ and peak-to-peak amplitude $A$, but the forward motion and flight speed are dynamically determined by the interaction with the fluid~\cite{becker2015hydrodynamic}. In the simulations, the single wing repeatedly traverses a domain of length $L$ with cyclic boundary conditions. Therefore, this setup is analogous to an infinite array of flyers with a fixed spacing between them.

\begin{figure}[htbp!]
    \centering
    \includegraphics[width=\textwidth]{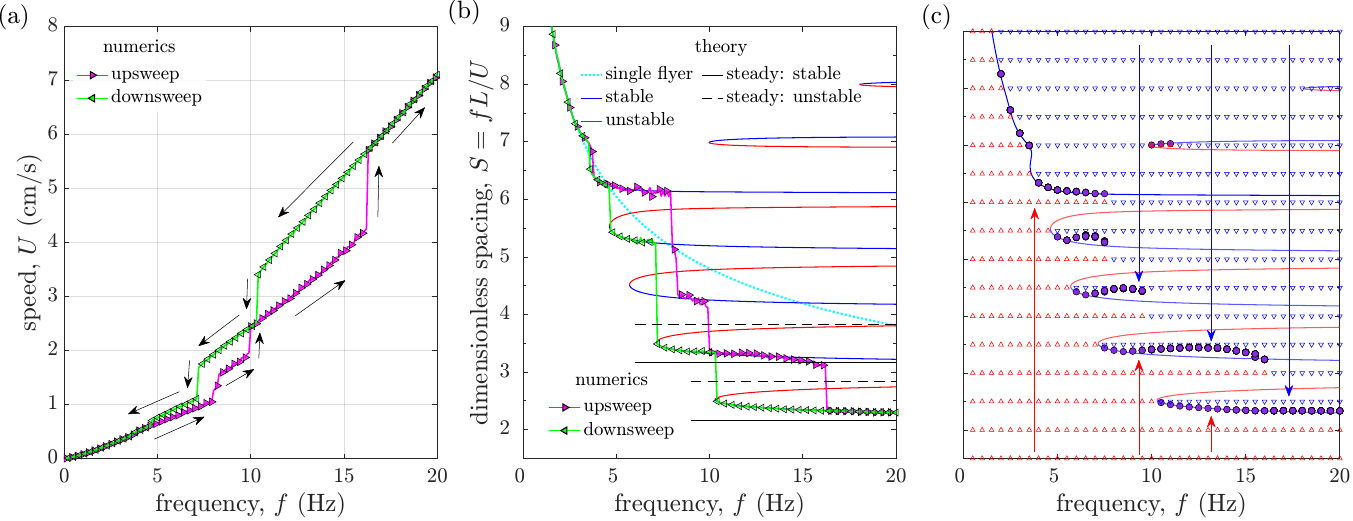}
    \caption{Propulsion dynamics of a self-interacting flyer. (a) Flight speed $U$ versus flapping frequency $f$, as a result of an upsweep of $f$ followed by a downsweep, as indicated by the black arrows. The simulations confirm the existence of multiple hysteresis loops. Here, $A=3$~cm. (b)~Same hysteresis loops as in (a), but in dimensionless spacing $S=fL/U$ versus~$f$ space. The dotted cyan line corresponds to a single, non-interacting flyer and it serves as a point of comparison. See Sec.~\ref{sec:equilibriumSpacing} for details. The solid blue and red curves represent the theoretical solutions, color-coded according to their stability: blue for stable and red for unstable solutions. The solid and dashed black lines in the high-$f$ and small-$S$ region correspond to $S=j+1/6$ and $S=j+5/6$ for $j\in\{2,3\}$, predicted as the stable and unstable $S$ values from the steady theory in Sec.~\ref{sec:stabilityAndStiffness}, when the wake dissipation timescale is assumed to be very slow ($\tau\to \infty$). (c) Equilibrium~$S$ values (purple circles) obtained numerically using different initial values $S_0$, indicated by the blue and red equispaced triangles, superposed with the theoretical $S$ values from panel~(b).}
    \label{fig:hysteresisA15}
\end{figure}

Experiments showed hysteresis loops corresponding to bistability of states: For identical flapping kinematics (same $A$ and $f$), there are slow and fast propulsion modes~\cite{becker2015hydrodynamic}. This bistability was also predicted using a discrete-time dynamical system that describes the hydrodynamic interactions between flyers arranged in one-dimensional lattice formations~\cite{oza2019lattices}.
Here, we vary $f$ and measure the resulting flight speed, while keeping $A$ fixed.
For a given frequency, we allow the system to reach a terminal flight speed and then we increase $f$ by a small amount, with this process repeated for other small increments of~$f$. This upward sweep (or, upsweep) is followed by a similar downward sweep (or, downsweep) to low flapping frequency values. 
The numerical results in Fig.~\ref{fig:hysteresisA15}(a) show that faster flapping through higher $f$, leads to faster propulsion, but the change in the flight speed $U$ is not always continuous. 
For example, $U$ increases relatively smoothly until a critical frequency, at which point it increases abruptly.
As $f$ is increased further,~$U$~again increases continuously until another abrupt increase is observed at $f\approx 10$ Hz. For decreasing $f$, the flight speed remains high before suddenly decreasing at $f\approx 16$ Hz. Similar to the upsweep process, further decreasing $f$ results in $U$ decreasing continuously, until another sudden decrease at~$f\approx 10$ Hz and so on. This implies that the system displays multiple hysteresis loops. We find that at low $f$, the upsweep and downsweep trajectories almost coincide, but increasingly higher flapping frequencies lead to larger gaps between the upsweep and downsweep trajectories, equivalent to wider hysteresis loops. 

To explain where and why transitions occur, we organize these observations in terms of the number of wavelengths separating successive flyers, by defining the dimensionless spacing $S$, computed using the flight forward speed~$U$ as $S=fL/U$. In Fig.~\ref{fig:hysteresisA15}(b) we show the same data as in Fig.~\ref{fig:hysteresisA15}(a), but in $f$-$S$ space. Since~$S$ is inversely proportional to the speed $U$, low frequency~$f$ results in large $S$, which corresponds to slow propulsion. This is indicative of weak flow interactions.
As~$f$ is increased, the numerically computed $S$, represented by the magenta and green triangles, decreases and plateaus near $j+1/6$ for $j\in\mathbb{N}$ for a broad range of frequencies. These values are the predicted stable $S$ values obtained from the steady theory in Sec.~\ref{sec:stabilityAndStiffness}, in the no-wake-decay limit ($\tau\to\infty$), which corresponds to long-lived wakes and is equivalent to strong interactions. The stable~$S$ values at $j+1/6$ with $j=2,3$ are shown as solid black lines in the high-$f$ and small-$S$ region, where the interactions are strongest.
Increasing $f$ induces abrupt downward jumps in $S$. Conversely, when $f$ is decreased, the system approaches $S\approx j+1/6$ before abruptly jumping upward to larger values of~$S$. This is precisely the hysteretic behavior associated with bistable locomotion states observed in Fig.~\ref{fig:hysteresisA15}(a). 

The follower-wake interaction model can be further utilized to determine equilibrium spacings for initial conditions that were not explored in experiments~\cite{becker2015hydrodynamic}.
A systematic search of equilibrium $S^*$ values, using different initial flight speeds (or, equivalently, initial spacings $S_0$) reveals multiple stable locomotion states for identical kinematic parameters.
This is evident from the coexistence of up to four stable $S^*$ values for the same frequency, as shown by the purple circles in Fig.~\ref{fig:hysteresisA15}(c). The red upward-pointing and blue downward-pointing triangles indicate the initial spacings $S_0$ that satisfy $S_0\leq \langle S^*\rangle$ and $S_0>\langle S^*\rangle $, respectively, where~$\langle S^*\rangle$ is the time-averaged dimensionless spacing in the equilibrium state, taken from the second half of the time series. 
This method of searching for multiple equilibrium states using an ensemble of initial conditions also successfully identifies stable $S^*$ at $S\approx 7+1/6$ with $f\approx 10$ Hz. 
These cases were not captured in experiments, which involved an upward sweep of frequency followed by a downward sweep to lower values.
Although the solid blue curves in Fig.~\ref{fig:hysteresisA15}(b), representing the stable branches of $S$ from theory, suggest that multiple stable locomotion states should exist for the same flapping kinematics, the upsweep and downsweep trajectories only fall on specific stable branches, avoiding others. Our systematic analysis in Fig.~\ref{fig:hysteresisA15}(c) confirms that certain predicted values of stable $S^*$ are actively avoided by the system. This behavior may be attributed to the inherent unsteadiness in the delayed system of equations, causing the self-interacting flyer to settle onto one of the other numerically computed $S^*$.

\subsection{Two-flyer problem with identical kinematics}\label{sec:ramananarivo}

\begin{figure}[htbp!]
    \centering
    \includegraphics[width=\textwidth]{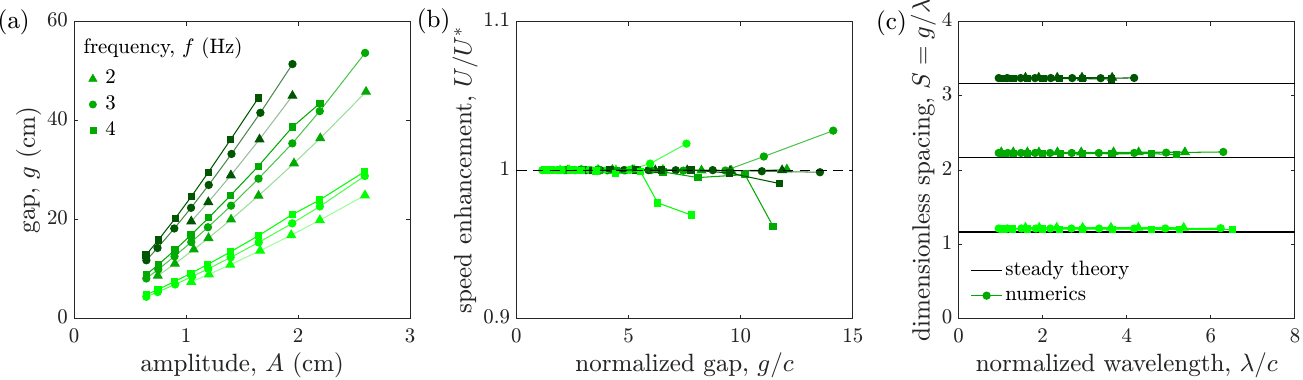}
    \caption{Emergent spacing and speed for a pair of synchronous flyers obtained by numerically solving Eqs.~\eqref{eq:xdot}--\eqref{eq:tnsystem} with cyclic boundary conditions. Darkening shades of green are used for increasing initial spacings. Frequency values of 2, 3, and 4 Hz are shown as triangle, circle, and square symbols, respectively. (a) Measured gap for varying values of $A$ and $f$. (b) Speed enhancement $U/U^*$ of a flyer pair relative to an isolated flyer, plotted against normalized gap, and (c) dimensionless spacing $S=g/\lambda$ as a function of the wavelength of the motion $\lambda$ normalized by the chord length $c$. The solid black lines in (c) represent the theoretical steady spacing values: $S=j+1/6$ for $j=1,2,3$. }
\label{fig:ramananarivofig2}
\end{figure}

The model is further validated using a two-flyer configuration, in which the two members are free to select both their speed and separation distance. See schematic in Fig.~\ref{fig:expValidationSchem}(c). 
Experiments have shown that when driven with identical flapping kinematics, the follower takes up one of several discrete, stable positions within the leader's wake, and the two travel together at the same speed~\cite{ramananarivo2016flow}.
In this case, the motions are characterized in terms of the steady-state equilibrium gap distance $g$ between the two flyers and the speed $U$ for different values of the kinematic parameters $A$ and~$f$, as shown in Figs.~\ref{fig:ramananarivofig2}(a) and~\ref{fig:ramananarivofig2}(b), respectively. The simulations demonstrate that different initial configurations yield distinct stable equilibrium arrangements that emerge from the flow interactions, consistent with the experiments~\cite{ramananarivo2016flow}. For each initial configuration, the separation distance initially changes before eventually reaching a steady-state value which thereafter remains constant. These steady-state values are all colored green, but with different levels of intensity: Darkening shades with increasing separation distance.
Figure~\ref{fig:ramananarivofig2}(a) illustrates that for any fixed flapping frequency~$f$, the gap $g$ in each state increases as the amplitude~$A$ increases. Increasing~$f$ also leads to greater values of $g$. 

A detailed comparison between the flight speed $U$ of the pair of flyers and the equilibrium speed~$U^*$ of a single, non-interacting flyer, as defined analytically in Eq.~\eqref{eq:Ueqm} and derived from steady theory in Sec.~\ref{sec:solo}, is shown in Fig.~\ref{fig:ramananarivofig2}(b), where the ratio $U/U^*$ is computed. 
This is equivalent to a speed enhancement factor, but since nearly all points lie on the line $U/U^*=1$ we deduce that the pair swims at approximately the same speed as an isolated flyer. Thus, the interactions have a negligible effect on $U$. While previous works revealed a modest speed enhancement, suggesting that the follower can influence the leader~\cite{ramananarivo2016flow,heydari2021school}, later studies found that flow interactions have negligible effects on  speed~\cite{newbolt2019flow,nitsche2024stability}, as in our model. The cyclic boundary conditions suggest that there might be a small influence of the follower on the leader, which may account for the small deviations from the $U/U^*=1$ line at large $g/c$ values in Fig.~\ref{fig:ramananarivofig2}(b), even at the $S\approx 3$ configuration (darkest green).

The emergent stable equilibrium states are better organized in terms of the dimensionless spacing $S=g/\lambda$, where $\lambda=U/f$ is the wavelength of the wave-like trajectory that the leader traces out as it moves through the fluid.
This is an alternative way of measuring the stable gap distance between the two flyers. The simulation data (green curves in Fig.~\ref{fig:ramananarivofig2}(c)) show that the pair falls on near-integer values of $S$, in agreement with experiments~\cite{ramananarivo2016flow}. The stable configurations at $S=j+1/6$ with $j=1,2,3$, derived from steady theory in Sec.~\ref{sec:equilibriumSpacing}, are represented by solid black lines in Fig.~\ref{fig:ramananarivofig2}(c). These results indicate that inter-flyer separations are quantized, with $(j+1/6)\lambda$ serving as the fundamental lattice spacing.

Experiments also revealed that the stability of the various positions assumed by the follower behind the leader depends on the strength of the hydrodynamic forces, which act to restore the follower's position when it is perturbed~\cite{ramananarivo2016flow}. 
This stability was mapped out by applying an external load to the follower, which would force it to adjust to a new spacing $S$ and a new steady speed $U$ for which the applied force would balance the net hydrodynamic force. 

\begin{figure}[htbp!]
\centering
 \includegraphics[width=\textwidth]{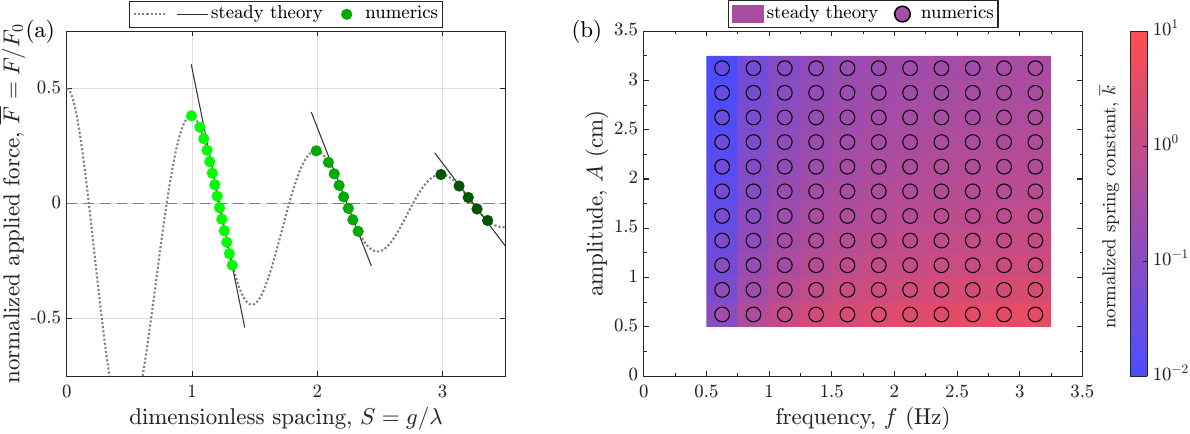}
    \caption{Restoring fluid force on the follower and springiness of the flow interactions. (a) Net fluid force $F$ on the follower versus dimensionless spacing $S=g/\lambda$ between the two flyers, for three initial spacings, represented by different shades of green (numerics), using cyclic boundary conditions. The kinematic parameters are: flapping amplitude $A=1.7$ cm and frequency $f = 3$ Hz. Here, $\lambda/c = 3.2$. The force $F$ is normalized by $F_0 = \rho C_Tcs(\pi  Af)^2/2$, which is the typical force due to dynamic pressure on a flyer with a planform area $cs = 29$ cm$^2$. The solid black lines represent the analytical form of $F=-k(S-S^{*})$ based on Hooke's law, where $k$ is given explicitly in Eq.~\eqref{eq:DeltaTSstarAndSpringConst}. The analytical form of $\overline{F}(S)$, defined in Eq.~\eqref{eq:eqmF}, is shown as a dotted black curve. (b) Spring constant $k$ derived from steady theory (colored background) and numerically computed $k$ (colored circles) in $f$-$A$ space, normalized by $F_0/c$ to give the dimensionless spring constant~$\overline{k}$. The results in~(b) correspond to an initial spacing $S_0\approx 1+1/6$. }
    \label{fig:ramananarivoDCforce}
\end{figure}

We incorporate the external steady forcing defined in Eq.~\eqref{eq:dcperturbation} into our model, and plot the force-spacing profile for the first three stable configurations as green shaded dots in Fig.~\ref{fig:ramananarivoDCforce}(a), with kinematic parameters $A=1.7$~cm and $f=3$~Hz. When no external force is applied on the follower ($F=0$), the first stable configuration (light green dot) corresponds to a dimensionless spacing of $S\approx 1+1/6$, which matches the theoretical $S^*$ from Sec.~\ref{sec:equilibriumSpacing}. If the follower is driven towards the leader ($F>0$), the spacing~$S$ decreases. Conversely, if the follower is driven away from the leader ($F<0$), $S$ increases and the system tries to restore the stable $S^*$ state. Therefore, the fluid force behaves as a Hookean spring, following $F\approx -k(S-S^*)$ near each stable state~$S^*$, where the spring constant $k$ is associated to the slope of the curves in Fig.~\ref{fig:ramananarivoDCforce}(a). 
The value of~$k$ depends monotonically on the equilibrium position of the follower:
A stiffer spring (larger~$k$) corresponds to flyers flying closer together. The first equilibrium is the most stable, as hydrodynamic interactions are strongest at shorter distances~\cite{ramananarivo2016flow,heydari2021school}.
This highlights the spring-like nature of the flow interactions, strengthening the view of the system as flow-mediated matter. 
We superimpose solid black lines representing the analytically predicted $F=-k(S-S^*)$, with~$k$ given by Eq.~\eqref{eq:DeltaTSstarAndSpringConst}. The negative slopes indicate that, for each equilibrium position, the hydrodynamic force acts as a restoring force that maintains the stability of the formation.

This analogy between the hydrodynamic interactions in the crystalline lattice and Hookean, one-way, diodic springs is further supported by additional characteristics, such as tensile and compressive strength. The latter corresponds to the maximum $\overline{F}$ that the lattice can withstand under compression before ``breaking.'' The model allows for a theoretical derivation of this value, which is a function of $S$, as discussed in detail in Sec.~\ref{sec:strengths} and shown in Fig.~\ref{fig:ramananarivoDCforce}(a). Larger $S$ values result in smaller maximum values of $\overline{F}$ before collisions occur between the flyers, leading to smaller compressive strengths.
We find that the system tolerates compressive strains up to approximately~20\%. This is determined by computing the relative difference between the strain $S$ at the maximum applied force $\overline{F}_{\max}$ which the bond can withstand, and the strain corresponding to the rest length of the spring, i.e.\ equilibrium spacing in the absence of any force. 
Similarly, the system can accommodate tensile strains of about $10\%$.
Beyond that, the system loses its elastic properties, akin to failure of a yield-stress material that deforms plastically.
We further note that, despite the applied forces, the speed of the pair of flyers~$U$ remains within 0.1\% of the speed of an isolated flyer $U^*$, suggesting that there is no speed enhancement from the force perturbation. However, we do not present the results here.

The spring constant associated with the flow interactions depends not only on the spacing between the two flyers, but also on the flapping parameters $A$ and $f$. While experiments reported results for three combinations of the kinematic parameters~\cite{ramananarivo2016flow}, we extend the analysis through simulations and theory to explore a broader range of $A$ and $f$ (still relevant to the experiments), as shown in Fig.~\ref{fig:ramananarivoDCforce}(b). Specifically, we present the analytical value of $k$ defined in Eq.~\eqref{eq:DeltaTSstarAndSpringConst}, normalized by $F_0/c$, as colored background in $f$-$A$ space. Additionally, the numerically estimated spring stiffness, obtained using the slopes of the best-fit lines from the $S$-$\overline{F}$ curves at the first equilibrium spacing, such as those in Fig.~\ref{fig:ramananarivoDCforce}(a), is overlaid as colored circles, with the color representing the normalized spring stiffness~$\overline{k}$.
The close match between the background colors and the colored circles for fixed $(f,A)$ combinations indicates strong agreement between the analytical and numerical values of~$\overline{k}$. Figure~\ref{fig:ramananarivoDCforce}(b) shows that the normalized spring stiffness $\overline{k}$ is positively correlated with~$f$ and negatively correlated with~$A$. In other words, the restoring fluid forces are strongest at low $A$ and high~$f$, and weakest at high~$A$ and low $f$.

\subsection{Two-flyer problem with distinct kinematics}\label{sec:newbolt2019}

Next, the model is validated using a two-flyer system, in which the leader and the follower have distinct flapping motions.
Specifically, the two actively flapping and passively interacting flyers can be synchronous or asynchronous, and can have dissimilar amplitudes and frequencies (see schematic in Fig.~\ref{fig:expValidationSchem}(d)). 
The vertical flapping velocity for the leader is given by $V_1= \pi A_1f_1\cos (2\pi f_1 t)$ and for the follower by ${V_2=\pi A_2f_2\cos(2\pi f_2t-\phi)}$. The following dimensionless kinematic parameters are varied: the initial follower phase lag $\phi\in[0,2\pi)$, the amplitude ratio $A_2/A_1$, and the frequency ratio $f_2/f_1$.
Since the leader's speed is determined by its own kinematics and is independent of the follower's motion, the dynamical states are characterized based on the follower's motion relative to the leader. Specifically, we use the dimensionless spacing $S=g/\lambda$, where $\lambda=U_1/f_1$~is the wavelength of the leader's flight trajectory and $g=X_1-X_2$ is the leader-follower gap distance at long times.

\begin{figure}[htbp!]
    \centering
 \includegraphics[width=\textwidth]{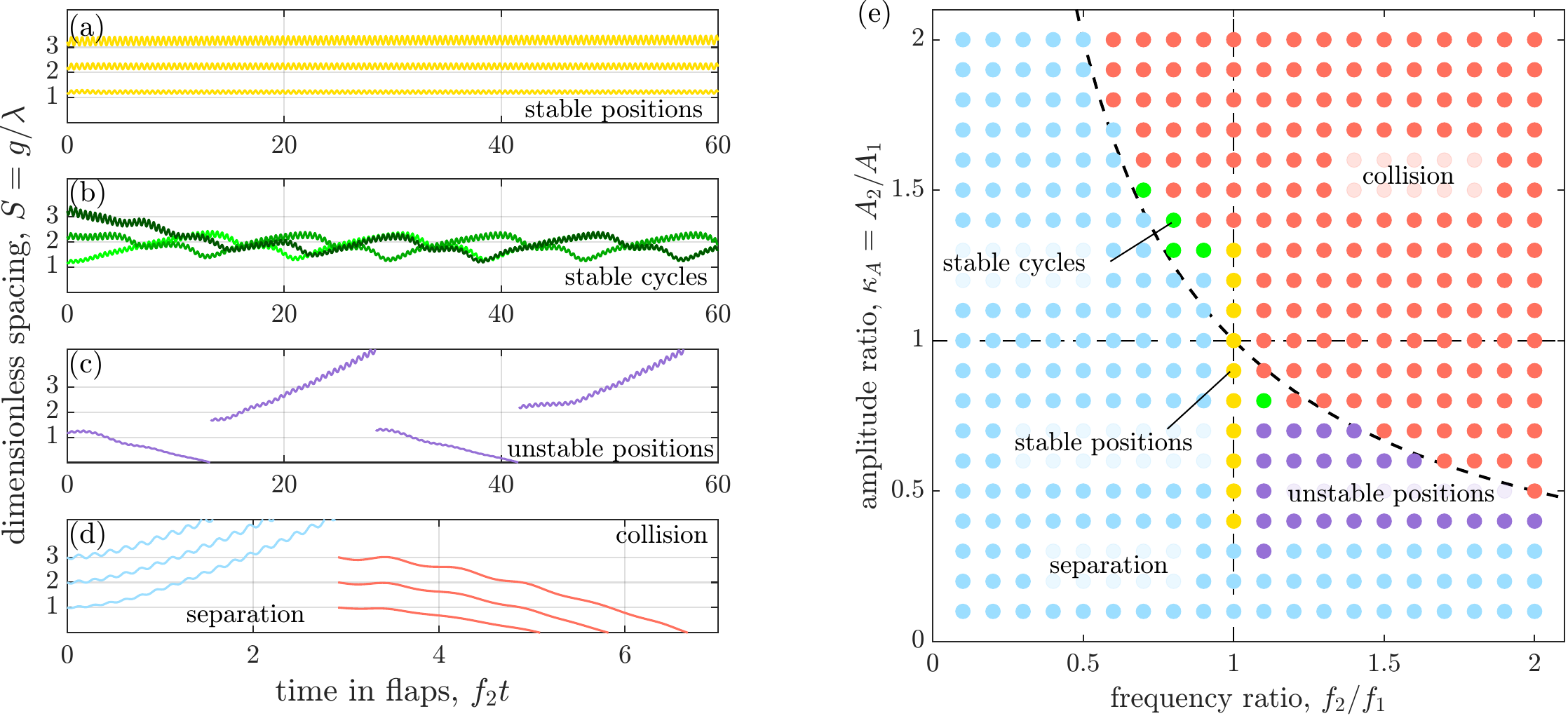}
    \caption{Variety of states achieved with different kinematic parameters and $\phi=0$. (a) The two flyers have identical flapping motions: $A_2/A_1=1$, $f_2/f_1=1$, and the follower takes up multiple stable positions. Here, $A_1=A_2=6$~cm and $f_1=f_2=2$~Hz. (b) An underdriven follower with $f_2/f_1<1$ and $A_2f_2/(A_1f_1)>1$ displaying stable cycles of $S$. Here, $f_2/f_1=0.9$ and $A_2f_2/(A_1f_1)=1.1$, with $f_2=2$ Hz and $A_2=8$ cm. (c) An overdriven follower with $f_2/f_1>1$ and $A_2f_2/(A_1f_1)<1$ has unstable positions in the form of either collisions into the leader or separations away from the leader, depending on the initial values of $S$. Here, $f_2/f_1=1.1$ and $A_2f_2/(A_1f_1)=0.8$, with $f_2 = 3.5$ Hz and $A_2=2$~cm. Each purple curve is the result of a different initial spacing. (d) The follower can also either always collide with the leader or always separate, regardless of the initial spacing. (e) Phase diagram for the five distinct states for two independently flapping flyers in the parameter space of $f_2/f_1$ and $A_2/A_1$. The follower's dynamical states are categorized by color: blue for separations, yellow for stable positions, green for stable cycles, purple for unstable positions, and red for collisions. The states with stable positions at $f_2/f_1=1$ divide regions where the follower has cyclic trajectories $f_2/f_1<1$, and unstable trajectories $f_2/f_1>1$.}
    \label{fig:fig4newbolt2019}
\end{figure}

Experiments have shown that a variety of states is achieved using different combinations of the kinematic parameters~\cite{newbolt2019flow}.
The simulation results, shown in Fig.~\ref{fig:fig4newbolt2019}, closely match these experimental observations. We begin by considering the simplest scenario, where the follower phase lag is zero ($\phi=0$) and the flapping amplitudes and frequencies of the two flyers are the same ($A_2/A_1=1$ and $f_2/f_1=1$). In this case, the flyers maintain a constant positive gap $g$ between them, corresponding to a stable positioning of the follower. The follower takes up one of several discrete preferred positions that have near integer values of dimensionless spacing $S$, as shown in Fig.~\ref{fig:fig4newbolt2019}(a), recovering the results presented in Sec.~\ref{sec:ramananarivo}. The dimensionless spacing tends to the steady-state equilibrium $S^*\approx j+1/6$ for $j\in\mathbb{N}$, derived theoretically in Sec.~\ref{sec:stabilityAndStiffness}. 
A time-varying, but bounded, $g>0$ implies stable cycling, as illustrated in Fig.~\ref{fig:fig4newbolt2019}(b).  
Here, we consider the situation of an overdriven follower, where $A_2f_2/(A_1f_1)>1$ and $f_2/f_1<1$. In this case, the phase $\phi$ is not important because the flapping phase changes continuously during the cycle, with the follower alternating between fast and slow propulsion.
In Fig.~\ref{fig:fig4newbolt2019}(c) we consider the opposite scenario, where the follower is underdriven with $A_2f_2/(A_1f_1)<1$ and $f_2/f_1>1$.
In this case, we observe multiple unstable positions for the follower, which either separates from or collides with the leader, depending on the initial value of $S$. 
If $g=0$ at any time, then a collision has occurred, and if $\dot{g}>0$, then the flyers are separating. This corresponds to a ``failed'' group. If the conditions $g=0$ or $\dot{g}>0$ are always achieved, regardless of the initial value of $S$, we classify these states as collision and separation, respectively, as shown in Fig.~\ref{fig:fig4newbolt2019}(d).

The dynamic behaviors of the flyer pair are categorized within the kinematic parameter space of $f_2/f_1$ and ${\kappa_A:=A_2/A_1}$ in Fig.~\ref{fig:fig4newbolt2019}(e), assuming a follower phase lag of zero. 
Different behaviors are distinguished by color, matching the colors used in panels (a)--(d). The special cases of $A_2/A_1=1$ and $f_2/f_1=1$, representing identical kinematics for the leader and follower, as well as $A_2f_2=A_1f_1$, are highlighted with dashed black lines. The latter curve marks the boundary between an underdriven follower (${A_2f_2/(A_1f_1)<1}$, below the boundary), and an overdriven follower (${A_2f_2/(A_1f_1)>1}$, above the boundary). The curve itself represents the condition of equal flapping speeds for the two flyers. One would expect the pair to separate when the follower is underdriven (blue dots), and indeed, this is true for most combinations of $A_2/A_1$ and $f_2/f_1$ and initial conditions. However, as shown in the case of $f_2/f_1=1$, this is not always the case. When the flapping frequencies are the same and $A_2/A_1\in[0.4,1.6]$, stable positions of the follower relative to the leader are observed (yellow dots). Under certain initial conditions (gap and speed difference), the follower may collide with the leader when $f_2/f_1>1$, even if it is underdriven; these cases are marked by purple dots in Fig.~\ref{fig:fig4newbolt2019}(d).
Conversely, one would expect the follower to collide with the leader (red dots) when it is overdriven. However, this is also not always the case. For some values of $f_2/f_1<1$, the follower displays stable cycles of $S$ behind the leader (green dots).
These findings are consistent with experiments~\cite{newbolt2019flow}.

\begin{figure}[htbp!]
    \centering
    \includegraphics[width=\textwidth]{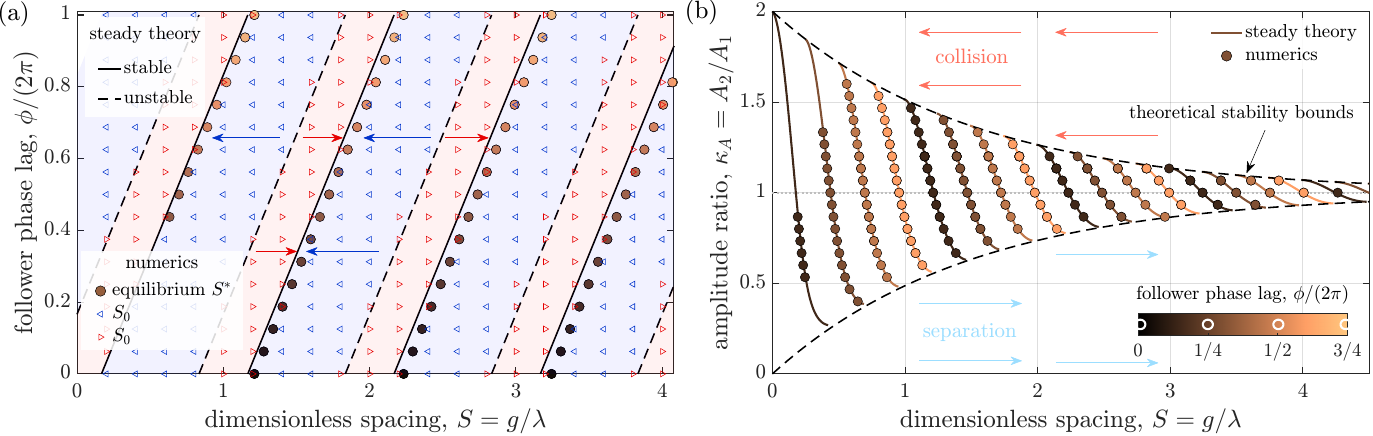}
     \caption{Stable positions for the follower achieved with different phase lags. (a) Stable equilibrium positions $S^*$ for the follower for varying~$\phi$ (numerics; brown-shaded circles), when $A_2/A_1=1$ and $f_2/f_1=1$. Increasing $\phi$ moves the stable positions downstream at a constant rate. If the follower is perturbed from one of the stable positions, it can fall into an adjacent stable position roughly one wavelength away. Here,  $A_1=A_2=4$ cm and $f_1=f_2=3$ Hz. The red and blue triangles indicate different initial spacings $S_0$, colored according to $S_0\leq \langle S^*\rangle$ and $S_0>\langle S^*\rangle$, respectively. The solid black lines denote the theoretical stable equilibria at $S^*=j+1/6+\phi/(2\pi)$, and the dashed black lines the unstable equilibria at $S^*=j-1/6+\phi/(2\pi)$. (b) Stable positions for the follower for varying amplitude ratios $\kappa_A=A_2/A_1$ and phase lags $\phi$, with $f_2/f_1=1$. The circles represent the numerical values of~$S$ and the solid brown-shaded curves satisfy Eq.~\eqref{eq:kappaAsoln}, obtained from steady theory.  
     Here, $A_1=4$~cm and $f_1=f_2=3$~Hz. The stable positions are enclosed between $\kappa_A=1-e^{-S/(f\tau)}$ and $\kappa_A=1+e^{-S/(f\tau)}$, shown by the dashed black lines, satisfying the inequality given by Eq.~\eqref{eq:kappaAinequality}. At fixed $\phi$, increasing $\kappa_A$ compresses the pair, while decreasing $\kappa_A$ leads to a dilated pair, up to some threshold after which the crystalline structure breaks down either through collisions or separations.}\label{fig:newbolt2019fig2}
\end{figure}

To focus on the effect of $\phi$ on the flyer dynamics, we restrict both the amplitude and frequency ratios to be equal to 1, and systematically vary the follower phase lag from 0 to~$2\pi$. For each value of $\phi$, we measure all stable positions and find that the follower assumes one of many discrete positions behind the leader, with the pair subsequently traveling together, recovering the results of experiments~\cite{newbolt2019flow}. This is illustrated in Fig.~\ref{fig:newbolt2019fig2}(a) in $\phi/(2\pi)$ versus~$S$ space. When $\phi=0$, the stable positions align with the results from previous sections, where the follower sits at a distance $S^*\approx j+1/6$, with $j\in\mathbb{N}$, from the leader. Increasing the follower phase lag $\phi$ leads to stable positions that are displaced downstream at a constant rate. In particular, the theory presented later in  Sec.~\ref{sec:stabilityKappaA} predicts stable positions with $S^*(\phi)\approx j+1/6+\phi/(2\pi)$. These are shown in Fig.~\ref{fig:newbolt2019fig2}(a) as solid black lines. The unstable values, $S^*(\phi)=j-1/6+\phi/(2\pi)$, are represented by the dashed black lines. As in the experiments~\cite{newbolt2019flow}, we show stable positions for $S\lesssim 4$; beyond this point, the flow interactions weaken~\cite{newbolt2019flow}. 
The model also enables us to efficiently populate the entire $S$-$\phi/(2\pi)$ plane with various initial configurations for the two-flyer system.
Specifically, in Fig.~\ref{fig:newbolt2019fig2}(a), we examine the nonlinear basins of attraction of the position equilibria by varying the initial separation distance $S_0$, and the phase lag $\phi$ between the two flyers. These basins of attraction are highlighted in red and blue, using as criteria: $S_0\leq\langle S^*\rangle$ and  $S_0>\langle S^*\rangle$, respectively, for increasing and decreasing $S$.
Therefore, this analysis shows that if the follower is perturbed from one of the stable positions, it can fall into an adjacent stable position approximately one wavelength away. This observation supports our interpretation of the flow interactions as flow-mediated matter, with the lattice spacing being about $1.2\lambda$.

If the frequency ratio is still fixed as 1, but both $\phi$ and the amplitude ratio $\kappa_A$ are varied, we find that the follower stably positions behind the leader for a wide range of these two dimensionless kinematic parameters.
The numerically computed stable positions $S$ for each~$\kappa_A$ are represented by circles in Fig.~\ref{fig:newbolt2019fig2}(b), colored according to the follower phase lag~$\phi$, with darker colors corresponding to smaller $\phi$. These coincide with the solid brown-shaded curves, which represent the theoretical stable positions for each $\phi$ and $\kappa_A$, given by Eq.~\eqref{eq:kappaAsoln} with the positive square root.
The stable region is bounded below by $\kappa_A=1-e^{-S/(f\tau)}$ and above by $\kappa_A=1+e^{-S/(f\tau)}$, illustrated as dashed black lines in Fig.~\ref{fig:newbolt2019fig2}(b). The derivation of these theoretical stability bounds can be found in Sec.~\ref{sec:stabilityKappaA}.
Within this stable region, the line $\kappa_A=1$ recovers the results in Fig.~\ref{fig:newbolt2019fig2}(a), where a larger $\phi$ results in a larger~$S$. This relationship holds for $\kappa_A\neq 1$ as well, although the stable states are observed over a narrower range of $S$ values. To better understand the physical implications of the results shown in Fig.~\ref{fig:newbolt2019fig2}(b), we consider a fixed phase lag. When $\kappa_A>1$, the gap distance between the leader and the follower is smaller than when $\kappa_A=1$, which means that increasing $\kappa_A$ compresses the pair. Instead, when $\kappa_A<1$, the spacing $S$ between them is larger than when $\kappa_A=1$, indicating that decreasing $\kappa_A$ effectively leads to a more dilated pair. Outside of the stable region, the two flyers no longer move cohesively. In particular, when $\kappa_A>1+e^{-S/(f\tau)}$ the follower is overdriven to an extent beyond what the pairwise hydrodynamic ``bond'' can sustain, which causes the follower to eventually collide with the leader. On the other hand, when $\kappa_A<1-e^{-S/(f\tau)}$ the follower is underdriven to a degree that leads to separation from the leader.

These findings further validate the effectiveness of the model and demonstrate its applicability to more complex setups. Viewing flying formations as flow-mediated matter, the kinematically distinguishable flyers can be seen as analogous to distinguishable particles in statistical mechanics, where each particle can be individually identified and tracked.

\subsection{Few-flyer problem}\label{sec:newbolt2022}

The detailed analysis of the two-flyer problem, with varying levels of complexity, provides ample evidence that the follower-wake model introduced in Sec.~\ref{sec:model} successfully reproduces the results of previous robophysical experiments~\cite{becker2015hydrodynamic,ramananarivo2016flow,newbolt2019flow}. Additional validation is provided by applying the model to the few-flyer problem, consisting of five flyers, as shown in the schematic diagram in Fig.~\ref{fig:expValidationSchem}(e).

\begin{figure}[htbp!]
    \centering
      \includegraphics[width=\textwidth]{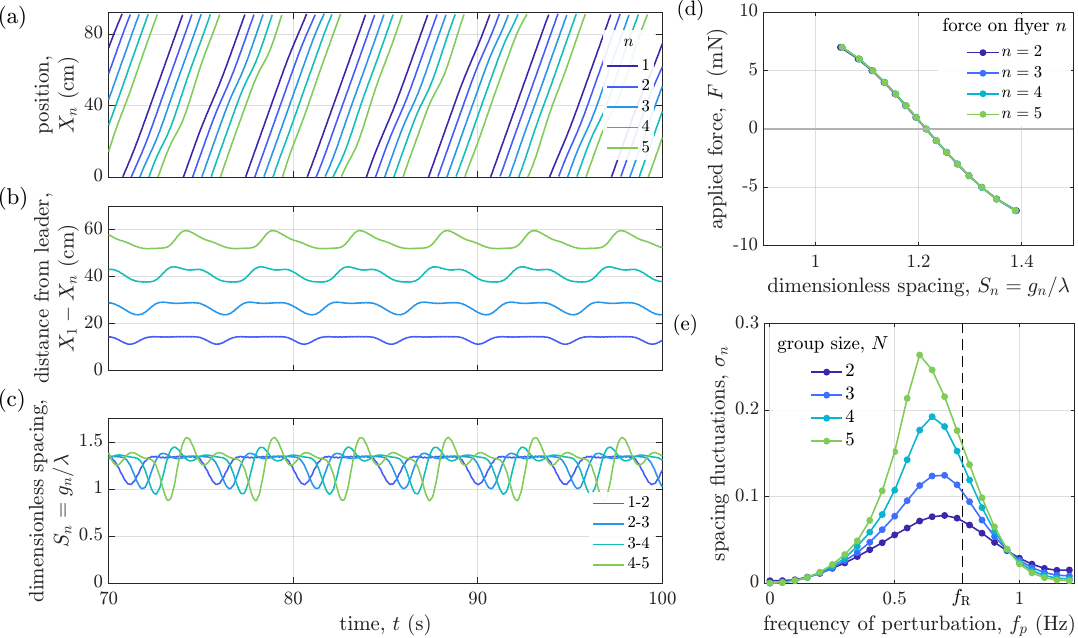}
    \caption{(a) Positions $X_n(t)$ for five flyers, using cyclic boundary conditions. (b) Positions of flyers $n=2,3,4,5$ relative to the leader $n=1$, showing strong fluctuations for later members. (c)~The dimensionless spacing $S_n=g_n/\lambda$ reveals that the members oscillate strongly in intermittent bursts. In (a)--(c), we focus on $t\in[70,100]$, where the motions are already in equilibrium, with kinematic parameters set to $A=3$ cm and $f=2.5$ Hz, and cyclic boundary conditions within a domain of length $C=92$~cm, as in experiments~\cite{newbolt2024flow}. (d) Interaction forces indicated by steady applied forces, defined in Eq.~\eqref{eq:dcperturbation}, on downstream members for a group of size $N=5$. (e)~Resonant oscillations in downstream members in response to  oscillatory forcing, as in Eq.~\eqref{eq:acperturbation}, applied on the leader. 
    In (d,e) the simulations are performed with open boundary conditions, such that the last member does not affect the leader, as in real in-line flying formations.}
    \label{fig:newbolt2022fig2}
\end{figure}

Experiments showed that followers tend to lock into stable positions behind a leader, but larger groups display flow-induced oscillatory modes, termed ``flonons,'' that grow in amplitude down the group and cause collisions~\cite{newbolt2024flow}.
The simulations confirm that the flying formation with few flyers is remarkably well ordered. Representative timeseries data for the positions $X_n(t)$ of members $n=1,2,\dots,N$ in a group of $N=5$ flyers are shown in Fig.~\ref{fig:newbolt2022fig2}(a). The lattice-like formation is demonstrated through the nearly uniform distances separating successive members in the group.
The crystal structure is also evident in Fig.~\ref{fig:newbolt2022fig2}(b), where we show the positions of each member in the group relative to the leader $n=1$. 
Positional fluctuations, which increase in amplitude down the group, are observed. These fluctuations are time-correlated and propagate through the group like traveling waves.
In Fig.~\ref{fig:newbolt2022fig2}(c) we track the gap distance between successive pairs  $g_n(t)=X_{n-1}(t)-X_n(t)$, normalized by the wavelength $\lambda$ of the undulatory trajectory of the leading member in each pair to give the dimensionless spacing $S_n=g_n/\lambda$. The time-averaged value of  $S_n$ is $\approx 1+1/6$; the stable $S^*$ from the steady theory analysis in Sec.~\ref{sec:stabilityAndStiffness}. Panel (c) indicates that the fluctuations have a temporal structure in the form of oscillations of $S_n$ that occur in sequence down the group and get stronger for members further downstream, exactly as in the experiments~\cite{newbolt2024flow}. As collective excitations that propagate in a lattice, these dynamics share some general features with conventional longitudinal displacement waves, such as phonons in atomic and molecular crystals~\cite{beatus2006phonons,chaikin1995principles}. However, the unique properties of the group motions, the nonreciprocal transmission and amplification, derive from the high Reynolds number flows.

Further insight into the emergent lattice formations and their dynamics can be gained by applying forces to individual members of the group. We begin by applying a steady force $F$ to one member in the group, as described by Eq.~\eqref{eq:dcperturbation}, similar to the two-flyer system in Sec.~\ref{sec:ramananarivo}. In a group of five flyers, we target the second flyer $n=2$. The simulations show that if the applied steady force is sufficiently small, the flyer will settle into a new position relative to its neighbors, and it will resume steady flight as the external  force is balanced by the fluid forces. Figure~\ref{fig:newbolt2022fig2}(d) shows force-displacement curves, generated for various magnitudes of the applied force and in both directions: towards and away from the upstream neighbor. The four force-displacement curves are colored from dark blue to green, based on whether the steady force is applied on the second, third, fourth, or fifth member, respectively. The curves behave similar to a spring that stabilizes the position with $S\approx 1+1/6$. The applied forces $F$ drive a given flyer towards its upstream neighbor and $S$ decreases. These are resisted by increasingly strong fluid forces of the same magnitude, until a maximum value is reached beyond which the flyer collides with its upstream neighbor. Similarly, a negative force $F$ results in a repositioning of the flyer to a larger spacing $S$ up to a point at which $F$ reaches a minimum and the flyer separates from its upstream neighbor. The high degree of overlap among the four curves (with $n=2,3,4,5$) suggests that the response differs to a small degree only across all members in the group that are not in the leading position, in agreement with experiments~\cite{newbolt2024flow}.
This is equivalent to the flow interactions being insensitive to the number of flyers upstream or downstream. 
The $S_n$-$F$ curves also agree well with previous measurements of the force on the follower in the 
two-flyer problem in Sec.~\ref{sec:ramananarivo}. We note that the nearest-neighbor property applies to the direct interactions mediated by flows, but more distant members still affect one another indirectly through the dynamics of intermediate members. Therefore, the group-wide dynamics are not immediately apparent from the pairwise interactions. All these results are consistent with experimental findings~\cite{newbolt2024flow}.

Experiments have also revealed that oscillatory perturbations applied to the leader induce resonant amplification in later members~\cite{newbolt2024flow}. Here, we apply an oscillatory force to the leader, given by Eq.~\eqref{eq:acperturbation}, and present in Fig.~\ref{fig:newbolt2022fig2}(e), the standard deviation $\sigma_n$ of the dimensionless spacing $S_n$ across a range of applied perturbation frequencies~$f_p$. 
The four curves are for the last member $n=N$ in groups of different sizes: $N=2$ (dark blue), $N=3$ (blue), $N=4$ (light blue), and $N=5$ (green). 
All members oscillate wildly near a resonance peak at $f_p\approx 0.6$ to 0.7~Hz. A slight shift to lower values of the resonant frequency is observed for downstream members, in agreement with experiments~\cite{newbolt2024flow}. Using steady theory, we also derive the analytical resonant frequency in Eq.~\eqref{eq:fRresonance}, as a function of the spring stiffness (Eq.~\eqref{eq:DeltaTSstarAndSpringConst}), which is associated to the slopes of the force-displacement curves in Fig.~\ref{fig:newbolt2022fig2}(d). The flapping frequency used to obtain the simulation results in Fig.~\ref{fig:newbolt2022fig2} is 2.5 Hz and the analytical resonant frequency in this case is $f_{\mathrm{R}}=0.77$~Hz (marked by the dashed vertical black line in Fig.~\ref{fig:newbolt2022fig2}(e)).
Downstream members experience amplified oscillations as can be seen from the larger spacing fluctuations $\sigma_5$, $\sigma_4$, and $\sigma_3$ compared to $\sigma_2$. This amplification property is the essence of flonons, observed in experiments and simulations, and is the main cause of the fragility of the crystalline groups of flyers.

This section shows that many of the results obtained for pairwise interactions carry over to the few-flyer system. Much like a lattice crystal with a small number of atoms, where the lattice spacing of $1.2\lambda$ is maintained and the atoms are held together by spring-like pairwise bonds, the few-flyer system also exhibits stable configurations. 
However, as the number of flyers increases, the system becomes more prone to instabilities, making the formation more fragile and susceptible to disruptions. These findings reinforce the validity of the follower-wake interaction model presented in Sec.~\ref{sec:model}, demonstrating its applicability to systems with more than two flyers.

\section{Model analysis}\label{sec:analytical}

To better understand the phenomena observed in previous experiments and in our simulations, we turn to theoretical analysis. By examining the two-flyer problem, we seek to explain how pairwise ``bonds'' form a larger crystalline structure and how hydrodynamic interactions in pairwise formations serve as a passive mechanism that promotes group cohesion.
In Secs.~\ref{sec:equilibriumSpacing}--\ref{sec:harmonic}, we present theoretical predictions for stable and unstable equilibrium spacings between two flyers with identical flapping motions.
We also provide an explanation of the springiness of the interactions and the tendency for disturbances to resonantly amplify. In Sec.~\ref{sec:stabilityKappaA}, we extend the theoretical analysis to flyers with dissimilar flapping amplitudes and a phase lag between their flapping motions, deriving the stable and unstable equilibrium spacings. 
Theoretical insights gained from the two-flyer system can be extended to the case of $N$ individuals. By studying these simpler interactions, we can extract analytical information that helps explain the behavior of the larger crystalline formation.
We further characterize some of its key properties through numerical simulations in~Sec.~\ref{sec:manybodies}.

\subsection{Equilibrium spacing and characteristic lattice constant}\label{sec:equilibriumSpacing}

To understand the origin of the crystallinity, we use the model equations and examine theoretically the two-flyer problem described in Sec.~\ref{sec:ramananarivo}, in which the two flyers have identical flapping motions: $A_2/A_1=1$, $f_2/f_1=1$, and $\phi=0$. 
Although the analysis here focuses on any leader-follower pair, the results carry over to the $N$-flyer case and, thus, the crystal in general. Below, we use steady theory to predict the stable and unstable equilibrium spacings, which correspond to the associated lattice constant.

The two flyers are flying in equilibrium if the distance between them and their time-averaged horizontal speeds over the flapping periods remain constant:
$g = X_1(t) - X_2(t)$ is a constant, and $\langle U_1\rangle_{T_f}=\langle U_2\rangle_{T_f}$, where $\langle\cdot \rangle_{T_f}$ denotes time-averaging over a flapping period $T_f=1/f$. The motion of the leader in the pair resembles that of a single flyer, whose equilibrium speed is $U^*$, as defined in Eq.~\eqref{eq:Ueqm}. The simulations presented in Fig.~\ref{fig:ramananarivofig2}(b) revealed that the flight speed of the pair is approximately the same as that of a single flyer.

In the absence of an external force, the follower's equation of motion is obtained by setting $n=2$ in Eq.~\eqref{eq:UndotDimensional}. By focusing on the thrust term and neglecting explicitly time-dependent terms (i.e., retaining only the ``DC terms''), we derive the following expression after some algebra:
\begin{equation}\label{eq:onlyThrust}
    \frac{\rho C_T cs}{2}(\pi Af)^2\left[\frac{1}{2}+\frac{1}{2}e^{-2\Delta t/\tau}-\cos(2\pi f\Delta t)e^{-\Delta t/\tau}\right].
\end{equation}
In Eq.~\eqref{eq:onlyThrust} the first term corresponds to the ``self thrust'' from non-interacting flapping, whereas the second and third terms correspond to the ``interaction thrust,'' which is directly related to the effect of the wake. As previously established, in equilibrium $\langle U_1\rangle_{T_f}=\langle U_2\rangle_{T_f}$, 
which requires that the interaction thrust is zero. This leads to the following transcendental condition for the steady equilibrium position $S^*$:
\begin{equation}\label{eq:transcendentalS}
    \cos(2\pi S^*)=\frac{1}{2}e^{ -S^*/(f\tau)}.
\end{equation}
To obtain this, we substitute $S^*=\langle U^*\rangle_{T_f} \Delta t/(\langle U^*\rangle_{T_f} /f)=f\Delta t$ into Eq.~\eqref{eq:onlyThrust}. Next, we consider the limiting case of \textit{no wake decay}, where the wake dissipation timescale is very slow ($\tau\to\infty$), resulting in strong flow interactions. In this limit ($f\tau\to\infty$), the transcendental condition for the steady equilibrium spacing, given by Eq.~\eqref{eq:transcendentalS}, becomes
    \begin{equation}\label{eq:theorySwithoutPhi}
        \cos(2\pi S^*)=1/2 \text{ and thus }S^{*}=j+1/6 \text{ or } j+5/6 \text{ for }  j\in\mathbb{N}.
    \end{equation}
This limiting case shows that between each pair of consecutive positive integers, there are two possible equilibrium spacings, corresponding to two distinct configurations. These are the possible arrangements in the systems discussed earlier in~Secs.~\ref{sec:becker} and~\ref{sec:ramananarivo}. 

The dimensionless equilibrium inter-member gap of about $1.2$ implies that the individuals tend to self-organize into a crystal-like formation with a characteristic lattice constant approximately equal to $1.2\lambda$. This is set by the wave-like wake flow which determines special positions where the thrust and drag forces come into balance. The time-averaged $S$-data in Figs.~\ref{fig:ramananarivofig2}(c) and \ref{fig:newbolt2022fig2}(c) confirm this analytical prediction, which applies not only for a leader-follower pair but also  more generally for $N\gg 1$.

We note here that for the single-flyer system with self-interactions, presented in Sec.~\ref{sec:becker}, the equilibrium solution $S^*$ at low flapping frequencies resembles that of an isolated, non-interacting flyer:
\begin{equation}
    S^*(f) = \left(\frac{\mu}{\rho f c}\right)^{1/3}\left(\frac{2C_D}{C_T}\right)^{2/3}\frac{L}{(\pi A)^{4/3}}.
\end{equation} 
This is determined through the equilibrium flight speed $U^*$ in Eq.~\eqref{eq:Ueqm} and is shown as a dotted cyan line in Fig.~\ref{fig:hysteresisA15}(b).
There, the $S^*(f)$ curve for the isolated flyer, in the interval $3\lesssim f\lesssim 9$, lies between the hysteresis loops corresponding to slow and fast modes. Physically, this means that the flow interactions can either be beneficial or destructive in terms of the flight speed.

\subsection{Stability of equilibrium spacings, spring stiffness, and resonance frequency}\label{sec:stabilityAndStiffness}
To analyze the stability of these equilibrium spacings, we introduce a spatial perturbation from equilibrium, $S=S^*+\Delta S$. This leads to a perturbation in thrust of the form:
\begin{align}\label{eq:DeltaT}
   \Delta T
   =\frac{\rho C_Tcs}{2}(\pi Af)^2&\left[\frac{1}{2}e^{-{2S^*}/{(f\tau)}}e^{-{2\Delta S}/{(f\tau)}}\right.\nonumber\\
   &\left.-[\cos(2\pi S^*)\cos(2\pi \Delta S)-\sin(2\pi S^*)\sin(2\pi\Delta S)]e^{-{S^*}/{(f\tau)}}e^{-{\Delta S}/{(f\tau)}} \right],
\end{align}
which is the interaction thrust, equal to the second and third terms in Eq.~\eqref{eq:onlyThrust}.
Since for $\Delta S$ small, $e^{-\Delta S/(f\tau)}\approx 1-\Delta S/(f\tau)$, $\cos(2\pi \Delta S)\approx 1$, and $\sin(2\pi \Delta S)\approx 2\pi \Delta S$, Eq.~\eqref{eq:DeltaT} simplifies to
\begin{equation}
    \Delta T=\frac{\rho C_Tcs}{2}(\pi Af)^2\left[\frac{1}{2}e^{-2S^*/(f\tau)}\left(1-\frac{2}{f\tau}\Delta S\right)-[\cos(2\pi S^*)-2\pi\sin(2\pi S^*) \Delta S]e^{-S^*/(f\tau)}\left(1-\frac{\Delta S}{f\tau}\right) \right].
\end{equation}
Using the transcendental condition in Eq.~\eqref{eq:transcendentalS}, the change in thrust force further simplifies to
\begin{equation}\label{eq:DeltaTSstarAndSpringConst}
    \Delta T=\hat{k}\Delta S=\frac{\hat{k}}{\lambda} \Delta g\quad ; \quad \hat{k}=\frac{\rho C_Tcs}{2}(\pi Af)^2\left[2\pi\sin(2\pi S^*)-\frac{1}{f\tau}\cos(2\pi S^*) \right]e^{-S^*/(f\tau)}.
\end{equation}
According to Hooke's Law, $F=k(g-g^{*})$, and so $k=\hat{k}/\lambda$ in Eq.~\eqref{eq:DeltaTSstarAndSpringConst} is analogous to the spring constant. 
This spring stiffness dictates the restoring action of the flow (see Fig.~\ref{fig:ramananarivoDCforce}(a)) and its stability.
Using Eq.~\eqref{eq:DeltaTSstarAndSpringConst} for the spring constant, we can also compute the resonant frequency $f_{\mathrm{R}}$, as follows:
\begin{equation}\label{eq:fRresonance}
    f_\mathrm{R}=\frac{1}{2\pi}\sqrt{\frac{k}{M}}=\frac{1}{2\pi}\sqrt{\frac{\hat{k}}{\lambda M}},
\end{equation}
where $M$ represents the mass and $\hat{k}$ is as in Eq.~\eqref{eq:DeltaTSstarAndSpringConst}. This theoretical resonant frequency is marked by a dashed vertical black line in Fig.~\ref{fig:newbolt2022fig2}(e). The spacing fluctuations $\sigma_n$ from the simulations indicate that the resonant frequencies are about 0.6 to 0.7 Hz, depending on the group size; with downstream members in larger groups shifting towards smaller values of~$f_{\mathrm{R}}$.
Note that, the analysis here focuses on a leader-follower pair, but since the results are applicable to the $N$-flyer problem, the form of Eq.~\eqref{eq:fRresonance} suggests that
the unstable growth of oscillations observed in Sec.~\ref{sec:newbolt2022} would be suppressed if members
have either different masses or different spring constants across individuals, as this would yield different resonant frequencies. Given the analytical form of $\hat{k}$ in Eq.~\eqref{eq:DeltaTSstarAndSpringConst}, the latter can be achieved either by introducing a vacancy defect into the crystal or when members have different flapping oscillations. This was confirmed experimentally in our earlier work~\cite{newbolt2024flow}.

For the equilibrium spacing $S^*$ to be stable, we require that $2\pi\sin(2\pi S^*)-\cos(2\pi S^*)/(f\tau)>0$, which ensures that $\Delta T$ and $\Delta S$ have the same sign, since the other scaling factors in Eq.~\eqref{eq:DeltaTSstarAndSpringConst} are positive. In particular, if $\Delta S>0$ (i.e., the follower is perturbed slightly away from the leader), then $\Delta T>0$ implies that the thrust acts as a restoring force, driving the follower back to its equilibrium position. 
This is equivalent to $S^*$ satisfying
\begin{equation}\label{eq:stabilityInequality}
    2\pi S^*\frac{\tan (2\pi S^*)}{\ln (1/(2\cos (2\pi S^*))}>1,
\end{equation}
which implies that $S^*= j+1/6$ is stable, while $S^*=j+5/6$ is unstable, where $j\in\mathbb{N}$ for both. This is corroborated by the simulation results in Figs.~\ref{fig:hysteresisA15}(b) and~\ref{fig:ramananarivofig2}(c).
Specifically, for the case of $N=1$ with self-interactions in Sec.~\ref{sec:becker}, we find that dimensionless spacings $S \approx j + 1/6$ for $j\in\mathbb{N}$ are favored by the system when the flyer is moving fast due to its high flapping frequency; a regime in which flow interactions are strongest. These theoretically stable states are represented by the solid black lines in Fig.~\ref{fig:hysteresisA15}(b). On the other hand, the unstable branches, typically occurring at $S\approx j +5/6$, are depicted by the dashed black lines and are avoided by the system.

\subsection{Compressive and tensile strength of inter-flyer bonds}\label{sec:strengths}

To evaluate the stability of the multiple equilibrium positions that a follower can adopt in the two-flyer system outlined in Sec.~\ref{sec:ramananarivo}, we applied an external steady force to the follower (see Fig.~\ref{fig:ramananarivoDCforce}(a)). Our results showed that the spring-like bonds between members provide elasticity to the group, allowing it to deform under loads and accommodate both compression and tension. However, the bonds possess a compressive and tensile strength beyond which the inter-flyer bond breaks. 
The maximum external load that the bond can endure without breaking can be derived analytically. We first assume that the pair is flying at equilibrium, meaning the gap between the two flyers remains constant: 
$X_1(t)-X_2(t)=g$. Taking a time-derivative of this, results in $\dot{X}_1(t)-\dot{X}_2(t)=0$, which implies that the horizontal speeds of the two flyers at equilibrium are equal: $U_1=U_2=U^*$.

To determine the strains that the system can tolerate, without losing its elastic response, we impose a steady force. With this perturbation on the right hand-side of Eq.~\eqref{eq:UndotDimensional} for $n=2$, and by retaining only the ``DC terms'' as in Eq.~\eqref{eq:onlyThrust}, we obtain
\begin{equation}\label{eq:withFeqm}
    \frac{\rho C_T cs}{2M}(\pi Af)^2\left(\frac{1}{2}+\frac{1}{2}e^{-2S/(f\tau)} -\cos(2\pi S) e^{-S/(f\tau)}\right)-\frac{C_Ds\sqrt{\rho\mu c}}{2M}(U^*)^{3/2}+\frac{F}{M}=0.
\end{equation} 

Substituting Eq.~\eqref{eq:Ueqm} for the equilibrium speed $U^*$ in Eq.~\eqref{eq:withFeqm}, and dividing throughout by the non-interactive thrust ${F_0 = \rho C_Tcs(\pi Af)^2/2}$, we obtain
\begin{equation}\label{eq:eqmF}
\overline{F} = \cos(2\pi S)e^{-S/(f\tau)}-\frac{1}{2}e^{-2S/(f\tau)}, 
\end{equation}
where $\overline{F}=F/F_0$ is the  dimensionless applied force. This $\overline{F}(S)$ is plotted in Fig.~\ref{fig:ramananarivoDCforce}(a) as a dotted black curve. Its extrema are determined by solving $\mathrm{d}\overline{F}/\mathrm{d}S=0$ with different initial guesses for $S$. Here, to compute the compressive strains we use as initial guesses 1, 2, and~3, which yield converged solutions $S_1^c$, $S_2^c$, and~$S_3^c$. Substituting these equilibrium spacing solutions into Eq.~\eqref{eq:eqmF} gives the compressive strength $\overline{F}(S_n^c)$ for each of the three initial configurations. These correspond to the peaks of the dotted black curve in Fig.~\ref{fig:ramananarivoDCforce}(a). Physically, this represents the maximum force the flyer-crystalline structure can withstand when compressed, without failing as a result of a collision.
Figure~\ref{fig:ramananarivoDCforce}(a) shows that the analytical prediction of the compressive strength (maxima of dotted black curves) matches the numerical results (green dots).

Similarly, to determine the tensile strains we use as initial guesses 1.5, 2.5, and~3.5, which yield converged solutions $S_1^t$, $S_2^t$, and $S_3^t$. Substituting these values into Eq.~\eqref{eq:eqmF}, we obtain the tensile strength $\overline{F}(S_n^t)$ of the crystalline formation, for each of the three initial arrangements. However, in this case, the theoretical predictions overestimate the numerical results shown in Fig.~\ref{fig:ramananarivoDCforce}(a) by about 40\%. The analytical predictions of the tensile strength in the different initial arrangements correspond to the troughs of the dotted black $\overline{F}$-curve. The theory also overestimates the tensile strains by about 60\%. This suggests that the unsteady terms in the governing equations (Eqs.~\eqref{eq:xdot}--\eqref{eq:tnsystem}), which are accounted for in the numerical calculations, reduce the material’s ability to withstand large tensile forces before losing its elastic response.

\subsection{Origin of flonons as resonantly amplifying waves}\label{sec:harmonic}
\begin{figure}[htbp!]
    \centering
    \vspace*{.25cm}
    \includegraphics[width=\textwidth]{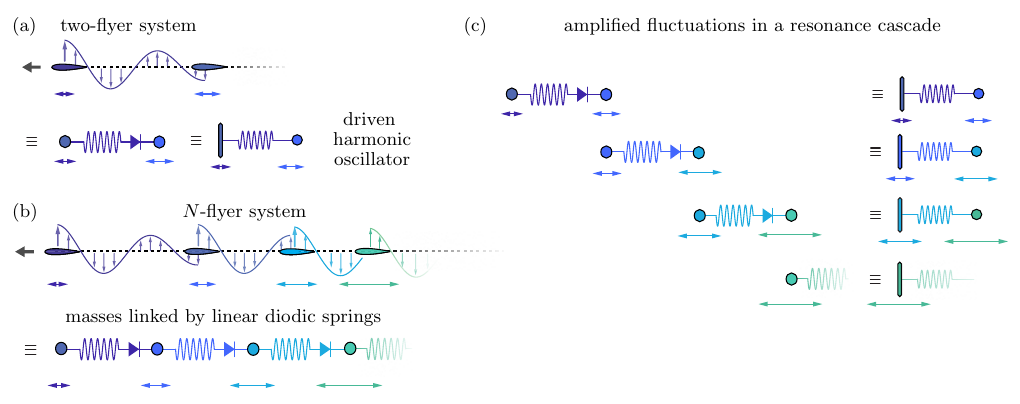}
    \caption{(a) Schematic diagram of a two-flyer system interacting through a wake emanated from the leader. This is analogous to a driven harmonic oscillator, where the leader plays the role of the driving source that forces the follower to oscillate. (b) The collective dynamics of $N$ flyers, each emitting a wake signal that gets erased by the downstream neighbor, viewed as a system consisting of a linear chain of masses linked to nearest neighbors by linear springs that are diodic. (c) Amplified fluctuations in a resonance cascade, where each pair of flyers can be isolated from the rest of the group given the one-way interactions.}
    \label{fig:springSchem}
\end{figure}

In our recent work~\cite{newbolt2024flow}, we demonstrated for the first time that a system of few flyers behaves similarly to a one-way, diodic mass-spring system, in the leader-to-follower sense. 
Our findings strongly suggest that the collective dynamics here is fundamentally different from conventional normal modes since the interactions are one-way~\cite{newbolt2024flow}.
The impact of this nonreciprocity on the amplification of positional fluctuations in the few-flyer system discussed in Sec.~\ref{sec:newbolt2022}, can be understood by considering a linear chain of masses linked to nearest neighbors by linear diodic springs, directed from the leader to the follower, as shown schematically in Fig.~\ref{fig:springSchem}(b).
To explain the mechanisms behind flonons (amplified fluctuations), we think of the leader in a pair of flyers as playing the role of the driving source that forces the follower to oscillate (see schematic in Fig.~\ref{fig:springSchem}(a)). This setup is analogous to a driven, damped harmonic oscillator.
The mechanisms behind amplified fluctuations in a resonance cascade for the few-flyer system are illustrated in Fig.~\ref{fig:springSchem}(c). We consider the response of the system to an oscillatory perturbation applied to the leader. The dynamics of the leader are fully determined, as the one-way interaction prevents flyer 2 (or any other downstream flyer) from influencing the leader's motion. According to classical results for a driven, damped harmonic oscillator, when damping is sufficiently weak and the perturbation frequencies are near the system's resonant frequency, the oscillation amplitude of flyer 2 is greater than that of flyer 1. This analysis makes use of the fact that pair 1-2 can be assessed as a system isolated from the rest of the group. This is because the 2-to-3 bond is a one-way interaction and therefore the dynamics of 3 do not influence those of 2. With the dynamics of 2 now completely determined, the next subsystem to be considered is 2-3. Similar to before, this can again be analyzed as isolated from all other downstream members. The same argument applies, and so 3 resonates even more strongly than 2. Iterating pairwise and sequentially down the group explains why the oscillations amplify~\cite{newbolt2024flow}. 
The amplitude ratio between successive members is the standard gain factor for the driven, damped harmonic oscillator, where member $n$ plays the role of the driving source that forces member $n + 1$ to oscillate, as shown schematically in Fig.~\ref{fig:springSchem}(c).

To systematically study this analogy between a driven, damped harmonic oscillator and the two-flyer system, we apply an oscillatory (AC) forcing to the leader, as described in Eq.~\eqref{eq:acperturbation}, with different driving frequencies. 
For each frequency, we compute the deviation of each flyer's perturbed position $X_n^{\mathrm{AC}}(t)$ from the unperturbed position~$X_n(t)$ --- the position in the absence of any external forcings. That is, ${\Delta X_n(t)= X_n^{\mathrm{AC}}(t)-X_n(t)}$. The gain factor is then defined as
\begin{equation}\label{eq:gainfactor}
  G=\frac{{\sigma}(\Delta X_{2}(t))}{{\sigma}(\Delta X_{1}(t))},
\end{equation}
where $\sigma(\Delta X_n(t))$ is the standard deviation of $\Delta X_n(t)$ at equilibrium and $n=1,2$.

\begin{figure}[htbp!]
    \centering
    \includegraphics[width=\textwidth]{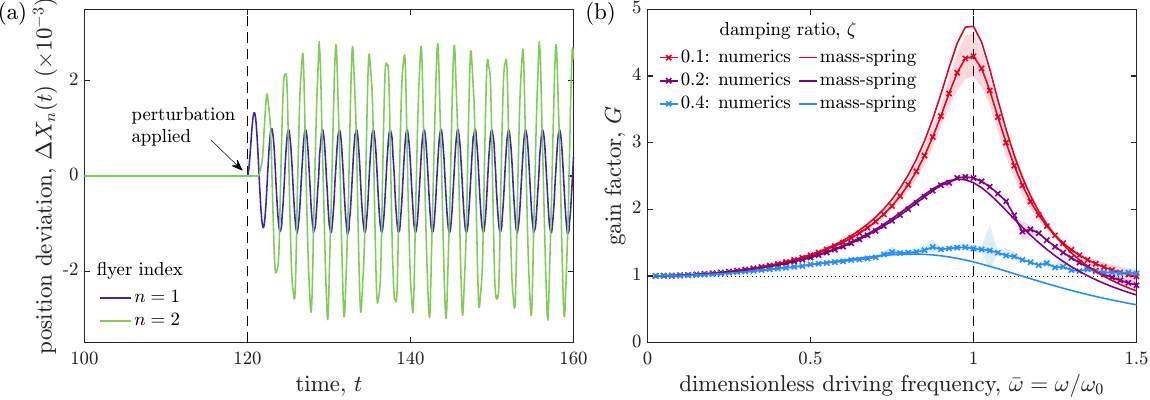}
    \caption{Quantification of amplification in the oscillation amplitude for the two-flyer system. (a) Example showing the difference between the perturbed position $X_n^{\mathrm{AC}}(t)$ and the unperturbed position $X_n(t)$ for the leader ($n=1$, in blue) and the follower ($n=2$, in green), when an external AC perturbation (as defined in Eq.~\eqref{eq:acperturbation}) is applied on the leader at the time indicated by the dashed vertical black line. The magnitude of the oscillations in $\Delta X_n(t)$ is determined by the amplitude of the perturbation force $a_{\mathrm{AC}}$ in Eq.~\eqref{eq:acperturbation}. By computing the standard deviation of each $\Delta X_n(t)$, the gain factor~$G$ in panel (b) can be calculated using Eq.~\eqref{eq:gainfactor}. (b) Resonance curves for the two-flyer system (solid lines with crosses), quantified by the gain factor as a function of the dimensionless driving frequency $\omega/\omega_0$. Each curve corresponds to a fixed dimensionless damping ratio $\zeta$ (as given in Eq.~\eqref{eq:zeta}), and is compared against the classical mass-spring system (solid lines without crosses). The shaded region encloses the maximum and minimum values of $G(\bar{\omega},\zeta)$ for different phases $\phi$ in~Eq.~\eqref{eq:acperturbation}.  Here, the AC perturbation amplitude is $a_{\mathrm{AC}}=10^{-3}M^*/(f^2A)$ and the frequency $f_{\mathrm{AC}}=\bar{\omega}\sqrt{k/M}/(2\pi f)$, with $A = 1.5$ cm, $f=2.5$ Hz, $M^*=0.16$, and $\bar{\omega}=\omega/\omega_0\in[0.025,1.5]$.}
    \label{fig:gainFactor}
\end{figure}

We present an example of how the gain factor is computed for a specific $\bar{\omega}=\omega/\omega_0$ value, in Fig.~\ref{fig:gainFactor}(a), where the driving frequency $\bar{\omega}$ appears in the AC forcing in Eq.~\eqref{eq:acperturbation} through $f_{\mathrm{AC}}=\bar{\omega}\sqrt{k/M}/(2\pi f)$. Specifically, we show $\Delta X_n(t)$ for the leader ($n=1$) in blue and for the follower ($n=2$) in green. The AC force, given by Eq.~\eqref{eq:acperturbation}, is applied at the time indicated by the dashed vertical black line. Prior to this, $\Delta X_n(t)=0$ for both $n=1$ and~2, but after the perturbation, the oscillation amplitudes of $\Delta X_2(t)$ are larger than those of $\Delta X_1(t)$, indicating that $G>1$ in this case.
We also show in Fig.~\ref{fig:gainFactor}(b) how the gain factor $G$ varies with the relative frequency $\omega/\omega_0$ for a set of damping ratios~$\zeta=0.1, 0.2, 0.4$. For the two-flyer system considered here, the damping ratio can be determined analytically using the change in the drag force resulting from a small perturbation in speed. Specifically, given $U=U^*+\Delta U$, the drag force becomes:
\begin{align}
   D=\frac{C_Ds\sqrt{\rho\mu c}}{2}(U^*+\Delta U)^{3/2}=\frac{C_Ds\sqrt{\rho\mu c}}{2}\left[(U^*)^{3/2}+\frac{3}{2}(U^*)^{1/2}\Delta U+\mathcal{O}\left((\Delta U)^2\right)\right]=D^*+\Delta D,
\end{align}
where $\Delta D=(3C_D/4)s\sqrt{\rho\mu cU^*}\Delta U$.
As with driven, damped harmonic oscillators, this can be expressed as ${\Delta D=\zeta\Delta U}$, where $\zeta$ represents the damping ratio:
\begin{equation}\label{eq:zeta}
    \zeta = \frac{3}{4}C_Ds\sqrt{\rho\mu cU^*}.
\end{equation}

For a fixed $\zeta$ and at each $\bar{\omega}$, we consider multiple phase lags between the leader's flapping motion and the oscillatory force. In Fig.~\ref{fig:gainFactor}(b), we shade the region between $\min_{\phi} G(\bar{\omega},\zeta)$ and $\max_{\phi} G(\bar{\omega},\zeta)$, using phase lags $\phi=0,\pi/4,\dots,7\pi/4$ in Eq.~\eqref{eq:acperturbation}. The average value of $G$ over all $\phi$ values is plotted as a solid line with crosses. Our simulations yield results that are qualitatively similar to those of the classical mass-spring problem (solid lines without crosses), particularly at small-to-moderate values of $\zeta$. As $\omega/\omega_0\to 0$, the perturbation becomes a constant force, and thus the velocity profile at equilibrium is identical for both flyers. This results in identical fluctuations in spacing, and consequently, the gain factor $G$ is exactly $1$. As the driving frequency increases, the effect of the perturbation becomes more pronounced, causing the spacing fluctuation of the second flyer to increase. For fixed $\zeta$, the gain factor reaches its maximum near resonance $\omega/\omega_0=1$, and then decreases, flattening out to saturation at $G\approx 1$. Therefore, similarly to Fig.~\ref{fig:newbolt2022fig2}(e), we find that when the leader is perturbed with an AC perturbation whose frequency is near resonance (and for sufficiently weak damping), the follower's oscillation amplitude will be larger than the leader's.

In the classical damped mass-spring system that is driven by sinusoidal motion of its support, the mass plays the role of the follower in a pair, and the support is the leader. Adopting standard notations, the equation of motion of the mass is $\ddot{x}+2\zeta\omega_0\dot{x}+\omega_0^2(x-x_0)=0$ that, with the driving $x_0(t) = A_0 \sin \omega t$,  becomes ${\ddot{x}+2\zeta\omega_0\dot{x}+\omega_0^2 x = \omega_0^2 A_0 \sin \omega t}$. Here $\omega_0 = \sqrt{k/m}$ is the resonance frequency, $c$ is the damping constant and $\zeta = c/(2m\omega_0)$ is its dimensionless form. The oscillatory forcing has amplitude $\omega_0^2 A_0$ and frequency $\omega$. This classical system has an associated gain factor $G = A/A_0 = \omega_0^2 / Z \omega$ with $Z = \sqrt{ (2 \omega_0 \zeta)^2 + (\omega_0^2-\omega^2)^2 / \omega^2}$, which can be cast in dimensionless form as
\begin{equation*}\label{eq:gain}
    G(\bar{\omega},\zeta) = \frac{1}{\bar{\omega}\sqrt{(2\zeta)^2 + (1-\bar{\omega}^2)^2 / \bar{\omega}^2}} \quad \rm where \quad \bar{\omega} = \frac{\omega}{\omega_0}.
\end{equation*}
We show resonance curves $G(\bar{\omega},\zeta)$ for three values of the damping ratio $\zeta$, as solid lines without crosses in Fig.~\ref{fig:gainFactor}(b). These results quantify that amplification occurs for sufficiently low damping and for driving frequencies $\omega$ near to or less than the resonance frequency~$\omega_0$.

In summary, the underlying instability is a two-body or nearest-neighbor effect. The oscillatory perturbations characterized in Fig.~\ref{fig:newbolt2022fig2}(e) of the current work and in experiments~\cite{newbolt2024flow} lead to amplification for groups of two to five flyers. However, resonantly amplifying waves appear for groups of all sizes. The main difference is that longer groups yield higher amplitudes, as displayed by the farthest downstream members. These large deviations from the lattice positions destabilize the formation and can lead to collisions between individuals.

\subsection{Equilibrium spacing and stability for dissimilar flapping amplitudes} \label{sec:stabilityKappaA}

A two-flyer system in which the leader and the follower flap with different amplitudes ($A_2/A_1\neq 1$) 
and have a phase lag between them ($\phi\neq 0$) is also amenable to theoretical analysis. However, with $f_2/f_1\neq 1$ analysis of the system becomes more prohibitive due to the complicated nonlinear dependence of the emergent dynamics on the flapping frequency. Therefore, in the remainder of this section, we fix $f_2/f_1=1$. 
The prescribed vertical velocities of the leader and the follower are given by $V_1(t)=\pi A_1f\cos(2\pi ft)$ and ${V_2(t)=\pi A_2f\cos(2\pi ft-\phi)}$, respectively. As in Sec.~\ref{sec:equilibriumSpacing}, the equilibrium requirement is ${\langle [V_2(t)-V_1(t-\Delta t)e^{-\Delta t/\tau}]^2 \rangle_{T_f} =\langle V_1(t)^2 \rangle_{T_f}}$, which gives
\begin{equation}
    \frac{A_2^2}{2}+\frac{A_1^2}{2}e^{-2\Delta t/\tau}-\langle 2A_1A_2\cos (2\pi ft-\phi)\cos (2\pi f(t-\Delta t))e^{-\Delta t/\tau} \rangle_{T_f} = \frac{A_1^2}{2}.
\end{equation}
The parameter $\kappa_A$ is employed to represent the ratio of flapping amplitudes ($\kappa_A=A_2/A_1$) and with the use of appropriate trigonometric identities and the usual definition of dimensionless spacing $S^{*}=f\Delta t$, we find that for a fixed $\kappa_A$, the condition for equilibrium spacing is now:
\begin{equation}\label{eq:quadraticKappaA}
    \kappa_A^2-2\kappa_A e^{-S^*/(f\tau)}\cos (2\pi S^*-\phi)+e^{-2S^*/(f\tau)}=1.
\end{equation}
Therefore, given a ratio of amplitudes $\kappa_A$, Eq.~\eqref{eq:quadraticKappaA} can be solved to obtain the equilibrium spacing $S^*$. Since Eq.~\eqref{eq:quadraticKappaA} is a quadratic equation for $\kappa_A$ there are two solutions associated with it. These are given explicitly by
\begin{equation}\label{eq:kappaAsoln}
    \kappa_A =e^{-S^*/(f\tau)}\cos (2\pi S^*-\phi)\pm \sqrt{1-e^{-2S^*/(f\tau)}[1-\cos^2(2\pi S^*-\phi)]}.
\end{equation}
Geometrically, the equilibrium condition in Eq.~\eqref{eq:quadraticKappaA} can be interpreted within the framework of a triangle, where it directly corresponds to the law of cosines. Therefore, the following inequality should be satisfied
\begin{equation}\label{eq:kappaAinequality}
    S^*\leq -f\tau \log |1-\kappa_A|.
\end{equation}
This inequality, upon rearrangement for $\kappa_A$ in terms of $S^*$, provides the theoretical lower and upper bounds for the stability region, shown in Fig.~\ref{fig:newbolt2019fig2}(b) as dashed black lines.

For the special case of $\kappa_A=1$ and $\phi=0$ in Eq.~\eqref{eq:quadraticKappaA}, which corresponds to a synchronous pair flapping with identical kinematics, we recover the transcendental equilibrium condition given in Eq.~\eqref{eq:transcendentalS}. For $\kappa_A=1$ and arbitrary $\phi\in[0,2\pi)$ we have $\cos(2\pi S^*-\phi)=e^{-S^*/(f\tau)}/2$
and we can again consider the limiting case of slow wake decay, $\tau\to\infty$. As $f\tau\to\infty$, the transcendental condition for the steady equilibrium spacing gives:
\begin{equation}\label{eq:theorySwithPhi}
    \cos (2\pi S^*-\phi)=1/2\text{ and thus }S^*=j \pm 1/6+\phi/(2\pi) \text{ for }j\in \mathbb{N}.
\end{equation}
These predicted equilibria are plotted in Fig.~\ref{fig:newbolt2019fig2}(a) as solid and dashed black lines for stable and unstable $S^*$, respectively.
Each of the intervals enclosing solutions dependent on the decay rate is shifted by $\phi$ (as a whole), but the $(S^*|_{\phi=0},S^*|_{\phi\neq 0})$ correspondence is not necessarily a linear shift by $\phi/(2\pi)$ for non-extreme values of~$\tau$.

For the stability criterion near the equilibrium spacing $S^*$, we consider a small perturbation about the equilibrium spacing $S^*+\Delta S$ and find that the stability criterion is now given by
\begin{equation}\label{eq:stabilityInequalitykappaA}
    \frac{2(\kappa_A^2-1)}{f\tau}+4\pi\kappa_Ae^{-S^*/(f\tau)}\sin (2\pi S^*-\phi)-\frac{2\kappa_A}{f\tau}e^{-S^*/(f\tau)}\cos(2\pi S^*-\phi)>0.
\end{equation}
This is the generalization of Eq.~\eqref{eq:stabilityInequality} for any $A_1$ and $A_2$ combination and for an arbitrary phase lag $\phi$.

These theoretical results reveal that followers can still lock in to the leader's pace even when their flapping kinematics are distinct. 
For instance, a weakly-flapping follower relocates to a downstream position in the wake of a faster-flapping leader, being pulled along behind it. Our findings determine the degree of variability among individuals that the material can tolerate.

\section{Implications for larger groups}\label{sec:manybodies}

By studying small, mesoscale groups composed of only a few members that are susceptible to collisions and structural fractures, as in the few-flyer system in Sec.~\ref{sec:newbolt2022}, we can gain valuable insights into the dynamics of longer in-line arrays.
While larger animal groups may display an underlying crystalline structure, they are inherently ``brittle,'' since the instabilities induced by the flapping motions are sufficient to trigger fractures in the group. An effective way of quantitatively analyzing large groups is to study how external perturbations affect the structural and dynamical properties of the formations. This allows us to measure key properties such as the degree of stability of the crystalline arrangements and the propagation speeds of these disturbances. Given the high nonlinearity of the many-flyer problem, we now return to simulations for its characterization.

\begin{figure}[htbp!]
    \centering
    \includegraphics[width=.61\textwidth]{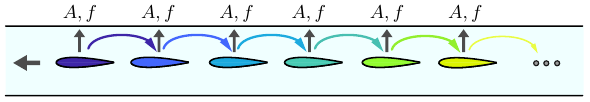}
    \caption{Schematic diagram of the arbitrary $N$-flyer problem, where all $N$ individuals have identical flapping motions ($A_i=A$ and $f_i=f$ for $i=1,\dots, N$). Open boundary conditions are applied in the simulations, such that the last member of the group does not affect the leader, thereby mimicking real in-line formations.}
    \label{fig:openSchem}
\end{figure}

In the preceding sections, the model is validated with previous experiments using cyclic boundary conditions and the dimensional form of the governing equations. To draw more general conclusions about in-line formations with an arbitrary number of flyers, here we use an open configuration with open boundary conditions, as illustrated schematically in Fig.~\ref{fig:openSchem}, and the  dimensionless system of equations~\eqref{eq:XndotDimensionless}--\eqref{eq:TndotDimensionless}, described in Sec.~\ref{sec:modelDimensionless}. In this case, the leader does not interact with the wake left behind by the last member in the group, and the wake dissipation constant $\tau$ tends to infinity. This corresponds to a very slow wake decay and, thus, strong interactions between the flyers. We drop the tildes for ease of notation, and seek to establish dimensionless groups of variables that can effectively describe the collective motion of a group of $N$ flyers. To this end, we focus on three key dimensionless groups derived in Eq.~\eqref{eq:4dimGroups}: $M^* = M / (\rho c^2 s)$, $\mathrm{Re}_f = \rho A f c / \mu$, and $A^* = A / c$. These quantities are particularly relevant, as they exhibit significant variation across different bird species flying in linear formations~\cite{greenewalt1962dimensional,viscor1987relationships,dimitriadis2015experimental,pennycuick1990predicting,dhawan1991bird,rayner1979new,krishnan2022role,taylor2019birds,maggini2017light,nafi2020aerodynamic,segretoexperimental}.

\subsection{Maximum cohesive group size}

Given the fragility of the group structure even with few flyers, we now raise the following question: What is the maximum number of flyers $N_{\max}$ that can fly cohesively in a linear formation, without colliding, by relying solely  on communication through the fluid flow, and in the absence of external perturbations? 
\begin{figure}[htbp!]
    \centering
    \includegraphics[width=\textwidth]{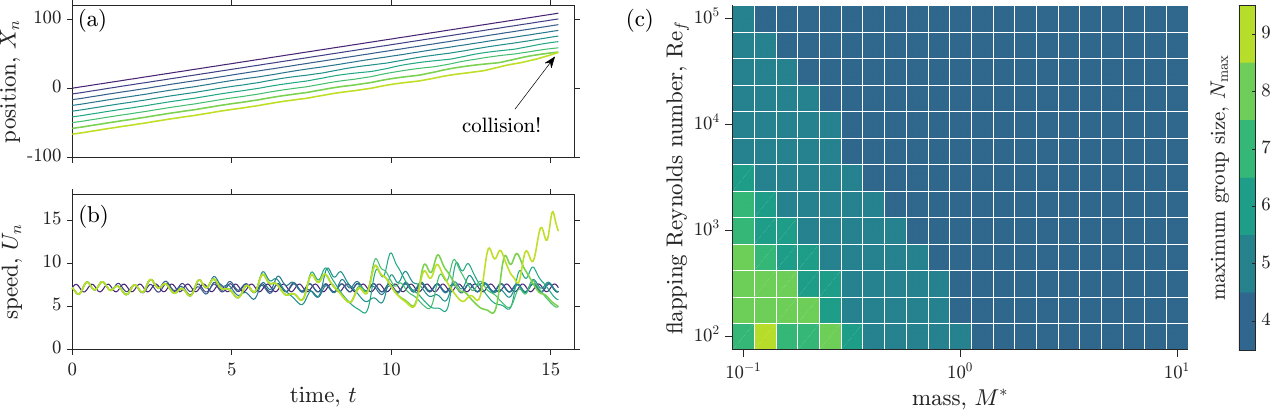}
    \caption{(a) Example of a collision between two individuals in a long array of flyers. Positions $X_n$ versus time for each flyer, computed up to the collision time. In this example, flyer 9 collides with its upstream neighbor, flyer 8. (b)~Flight speed $U_n$ versus time for each of the nine flyers in the same example. (c) Maximum group size $N_{{\max}}$ for flyers traveling cohesively without collisions in $M^*$-$\mathrm{Re}_f$ space, as colored background. Here, $A^*=0.25$.}
    \label{fig:collisionExample}
\end{figure}

We start by presenting an example of a collision between two flyers within a group, in the absence of external disturbances. The initial speed of each flyer is set equal to the steady speed of the group leader, Eq.~\eqref{eq:Ueqm}, which means that the flyers are initialized slightly away from their equilibrium positions and speeds. This is because the equilibrium flight speed is not constant over time, as seen in Fig.~\ref{fig:collisionExample}(b). As a result, this small initial deviation from equilibrium can sometimes lead to a collision. 
To determine through simulations whether a collision occurs, we define a gap-threshold between adjacent flyers that is slightly greater than zero. If this threshold is exceeded, the numerical integration terminates. Figures~\ref{fig:collisionExample}(a) and \ref{fig:collisionExample}(b) illustrate this process, showing the positions and speeds of the flyers over time. In this case, flyer 9 collides with its upstream neighbor, flyer 8.

Based on this kind of data, we determine $N_{\max}$ for different combinations of the three underlying dimensionless groups: mass $M^*$, flapping Reynolds number $\mathrm{Re}_f$, and flapping oscillation amplitude $A^*$. 
We present as colored background in Fig.~\ref{fig:collisionExample}(c), the maximum size $N_{\max}$ of a linear chain of flyers, in $M^*$-$\mathrm{Re}_f$ space, with $A^*$ fixed as 0.25.
The simulations reveal that, for the range of dimensionless parameters considered in this study, the crystalline structure is maintained without fracture for a maximum of 4 to 9 flyers. This is consistent with the analysis of the two-flyer system, which showed that disturbances can resonantly amplify, severely limiting the crystal's size. The largest stable groups are found in the lower left region, where both $M^*$ and $\mathrm{Re}_f$ are small. For small-to-moderate values of $M^*$, $N_{\max}$ decreases monotonically with increasing $\mathrm{Re}_f$, and for $M^*> 1$ only groups consisting of 4 flyers remain stable. This highlights the brittleness of the crystal, since even small perturbations can lead to unstable growth and collisions between flyers.

\subsection{Instability growth timescale and disturbance propagation speed}

\begin{figure}[htbp!]
    \centering
    \includegraphics[width=\textwidth]{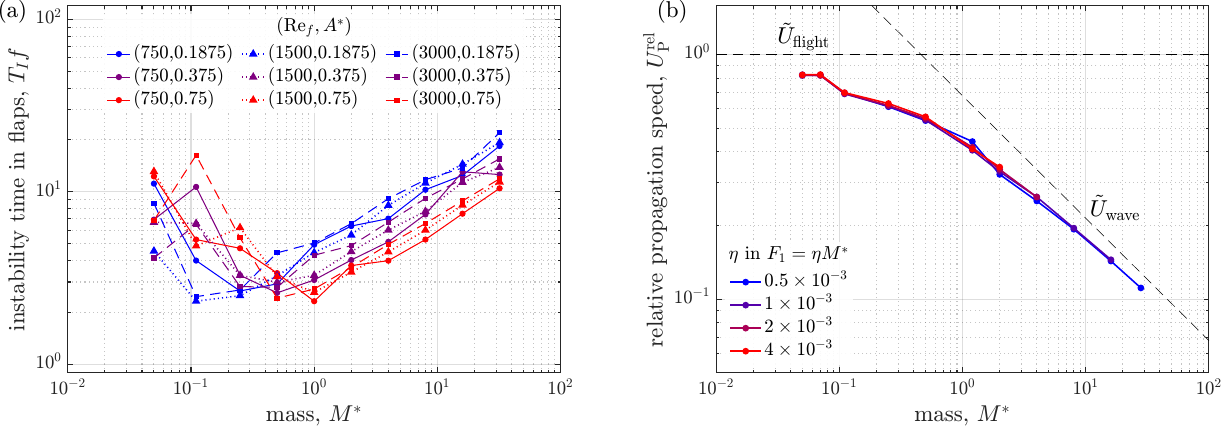}
    \caption{(a) Instability time (from initialization to collision), measured in flaps and calculated as $T_{I}/T_f=T_If$, versus dimensionless mass $M^*$ for various $(\mathrm{Re}_f,A^*)$ combinations. 
    (b) Relative propagation speed of disturbance to flight speed $U_{\mathrm{P}}^{\mathrm{rel}}$ versus  $M^*$. The horizontal dashed line denoted by $\tilde{U}_{\mathrm{flight}}$, at $U_{\mathrm{P}}^{\mathrm{rel}}=1$, is relevant to the small-$M^*$ limit, where the disturbance propagation speed is limited by the flight speed. The slanted dashed line denoted by $\tilde{U}_{\mathrm{wave}}$ is relevant to the large-$M^*$ limit, and is the group speed of a wave propagating in a diodic mass-spring system, given in Eq.~\eqref{eq:uwave}, appropriately non-dimensionalized. These results are obtained for a group of $N=3$ flyers with $(\mathrm{Re}_f,A^*)=(400,0.25)$ and $F_1=\eta M^*$ in Eq.~\eqref{eq:ImpulseForceGaussian}, where ${\eta \in\{0.5,1,2,4 \}\times 10^{-3}}$.}
    \label{fig:instabilityTimescale}
\end{figure}

The same kind of data and, in particular, the flight speed $U_N$ of the flyer that collides with its upstream neighbor (see, for example, the light green curve in Fig.~\ref{fig:collisionExample}(b)), can be used to compute the timescale of instability growth for different combinations of $M^*$, $\mathrm{Re}_f$, and $A^*$. We define $\Delta U_N:=U_N-U^*$, which represents the deviation of the flight speed of flyer $N$ from the theoretical equilibrium speed given in Eq.~\eqref{eq:Ueqm}. The slopes of $\ln|\Delta U_N|$ give the growth rates. An instability starts to develop at $t=t_0$ with growth rate~$\gamma$, its form given by $\Delta U_N\sim e^{\gamma(t-t_0)}$. Using this, the timescale of instability growth $T_I$ can be computed as~$1/\gamma$.
We re-cast $T_I$ in terms of the number of flaps needed to avoid collisions, a language that offers physical intuition and lends itself to biological interpretation.
This is obtained by computing $T_{I}/T_f=T_If$, where $T_f=1/f$ is the flapping period and is equal to $1$ in its dimensionless form; see Eq.~\eqref{eq:tildeQuantities}. In Fig.~\ref{fig:instabilityTimescale}(a) we show the instability time in flaps for a range of values of $M^*\in[0.05,32]$, using nine combinations of $(\mathrm{Re}_f,A^*)$, including the baseline case $(\mathrm{Re}_f,A^*)=(1500, 0.375)$ used in experiments~\cite{newbolt2024flow}. We find that for all values of $M^*$ here, the flyers have to react very quickly, within about 2 to 20 flaps, to avoid collisions. 
The instability time in flaps is $\mathcal{O}(10)$ in all cases, but there are two distinct limits of $M^*$ (small and large) where the flyers are generally slightly more ``safe.'' Flyers with moderate values of $M^*$ ($\approx 1$), have only $2\sim 5$ flaps before they collide with their upstream neighbor. We note that this is also the critical value of $M^*$, which distinguishes between groups of 4 flyers and slightly larger groups in Fig.~\ref{fig:collisionExample}(c). This analysis suggests that physical interactions are important in setting the gross structure of flying formations and that sensing-and-response is needed to maintain or stabilize them.

Another intriguing analogy to the classical mass-spring system is the propagation speed of a disturbance as it travels down the group. This can be computed by applying an impulse force $F_I(t)$ to the leader in a group of three flyers, as described in Eq.~\eqref{eq:ImpulseForceGaussian}. The choice of such a small group is deliberate, as we find that, even in the absence of external perturbations, the maximum number of flyers for a stable formation to exist at large values of $M^*$ is four (see Fig.~\ref{fig:collisionExample}(c)). The relative propagation speed of disturbance to flight speed is defined as
\begin{equation}\label{eq:Unprop}
U_{\mathrm{P}}^{\mathrm{rel}}=\frac{|U_{\mathrm{P}}-\overline{\langle U_{n}\rangle}|}{|\overline{\langle U_{n}\rangle}|},    
\end{equation}
where $U_{\mathrm{P}}$ and $\langle U_{n}\rangle$ are the propagation speed of the disturbance in the lab frame and the time-averaged flight speed of flyer $n$, respectively. We note that at equilibrium, $U_n$ is approximately the same for all flyers $n=1,2,3$ and so in calculating $\overline{\langle U_n\rangle}$ we take the average of $\langle U_n\rangle$ among all flyers. Equation~\eqref{eq:Unprop} can be viewed as a transformation of the lab frame propagation speed to a group frame propagation speed. 
We show $U_{\mathrm{P}}^{\mathrm{rel}}$ versus $M^*\in [0.05, 28]$ on a log-log scale in Fig.~\ref{fig:instabilityTimescale}(b), for four different impulse force magnitudes $F_1=\eta M^*$ in Eq.~\eqref{eq:ImpulseForceGaussian}, where $\eta \in\{0.5,1,2,4 \}\times 10^{-3}$. Here, $\mathrm{Re}_f$ and $A^*$ are fixed as 400 and $0.25$, respectively. Stronger impulse forces lead to collisions, so we do not pursue those here. 
Figure~\ref{fig:instabilityTimescale}(b) shows that $U_{\mathrm{P}}^{\mathrm{rel}}$ decreases with~$M^*$, indicating that greater inertia of the flyers leads to slower propagation of disturbances.

We suggest an interpretation of these data that involves two regimes of wave propagation and their respective speeds. One contributing effect involves the fundamental delay associated with flow disturbances being left by one member and picked up later by the downstream neighbor, and the second effect pertains to elastic waves mediated by the nonreciprocal spring-like bonds. 
The former operates irrespective of mass and represents the upper limit to the speed. 
Since the wake flow has no horizontal propagation, information can only be transmitted as fast as the flight speed $U_{\mathrm{flight}}$, corresponding to the horizontal dashed line in Fig.~\ref{fig:instabilityTimescale}(b). Comparison to the data show that this effect is the primary factor for low mass $M^*$. Higher $M^*$ imposes additional limits associated with sluggishness of heavy flyers to elastically respond, and this mass-dependent effect is shown as the sloped line that results from the calculation described below for the wave speed $U_{\mathrm{wave}}$. 

We model our calculations after recent work that considered the continuum analog of nonreciprocal elasticity~\cite{brandenbourger2019non}, the results of which are extended here for a discrete system of masses connected by one-way or diodic springs. 
The equation of motion for mass $n$ is $M\ddot{u}_n+k(u_n-u_{n-1})=0$, with $n=2,3,\dots, N$. Assuming plane wave solutions of the form $u_n\propto e^{i(\Omega t-q_m n\lambda)}$, the dimensional form of the complex-valued dispersion relation is given by
\begin{equation}\label{eq:dispersionNonreciprocal}
    \Omega(q_m)=\sqrt{\frac{k}{M}}\left(1-e^{iq_m\lambda}\right)^{1/2},
\end{equation}
where $\Omega$ is the wave angular frequency, $k=\hat{k}/\lambda$ is the spring stiffness defined in Eq.~\eqref{eq:DeltaTSstarAndSpringConst}, $\lambda$ is the lattice spacing, and $q_m$ is the wave number for mode $m$. 
Differentiating the real part of $\Omega(q_m)$ with respect to $q_m$ gives the group speed for longitudinal waves:
\begin{equation}\label{eq:uwave}
    U_{\mathrm{wave}}(q_m) = \frac{\d }{\d q_m}[\Re\{\Omega(q_m)\}] = \frac{\lambda}{2r}\sqrt{\frac{k}{M}}\,\,[\sin(q_m\lambda)x-\cos(q_m\lambda) y],
\end{equation}
where $r=\sqrt{\alpha^2+\beta^2}=\sqrt{2\alpha}$, $x=\sqrt{r}\cos[(\arctan\left(\beta/\alpha \right))/2]$, ${y=\sqrt{r}\sin[(\arctan\left(\beta/\alpha \right))/2]}$, ${\alpha=1-\cos(q_m\lambda)}$ and $\beta=-\sin(q_m\lambda)$. The eigenmodes are combinations of $\cos(q_m n\lambda)$ and $\sin(q_m n\lambda)$ satisfying fixed--free boundary conditions given by $u_{1}=0$ and $u_{N-1}=u_{N+1}$, respectively. The latter condition arises from central finite difference discretization, after introducing a ghost point at $n=N+1$ to ensure a free boundary condition at $n=N$ (i.e.\ $u'=0$ at $n=N$).
The allowed wave numbers $q_m$ correspond to values for which the determinant of the matrix 
\begin{equation}
    \begin{pmatrix}
    \cos(q_m\lambda)&\sin(q_m\lambda)\\
    \cos[q_m(N+1)\lambda]-\cos[q_m(N-1)\lambda]&\sin[q_m(N+1)\lambda]-\sin[q_m(N-1)\lambda]\\
    \end{pmatrix}
\end{equation}
vanishes. The possibilities include $q_m=m\pi/\lambda$ or $q_m=(2m\pm 1/2)\pi/[(N-1)\lambda]$ with $m\in\mathbb{Z}$, and $q_m$ is restricted to the
first Brillouin zone $(-\pi/\lambda,\pi/\lambda]$. 
The physically relevant wave numbers are those for which the wave propagates in the downstream direction, i.e. $q_m>0$. For the case of $N=3$ flyers considered in Fig.~\ref{fig:instabilityTimescale}(b), the possible $q_m$ are therefore $\pi/\lambda$, $3\pi/(4\lambda)$, and $\pi/(4\lambda)$. The disturbance propagation speed $U_{\mathrm{P}}^{\mathrm{rel}}$ is bounded above by the fastest wave speed $U_{\mathrm{wave}}$ (in Eq.~\eqref{eq:uwave}), which occurs when the wave number is equal to $\pi/(4\lambda)$. After appropriate non-dimensionalization, we arrive at the curve for $U_{\mathrm{wave}}$ displayed as the dashed slanted line in Fig.~\ref{fig:instabilityTimescale}(b). Comparison to the data shows that this effect provides the limit to the wave speed in the limit of high mass. 

The physical interpretation of why $U_{\mathrm{P}}^{\mathrm{rel}}$ decreases with mass is that each flyer slows down slightly later than the previous one, resulting in a disturbance propagation speed that is slower than the flight speed, with the disturbance effectively moving forward in the lab frame. The inertia of the flyers plays a significant role in this behavior. Larger mass means that the flyer takes longer to slow down after encountering the disturbance, causing each flyer to leave the disturbance in its wake slightly ahead of where it first encountered the perturbation.

\subsection{Fragility of the group assessed by force perturbations}

\begin{figure}[htbp!]
    \centering
    \includegraphics[width=.98\textwidth]{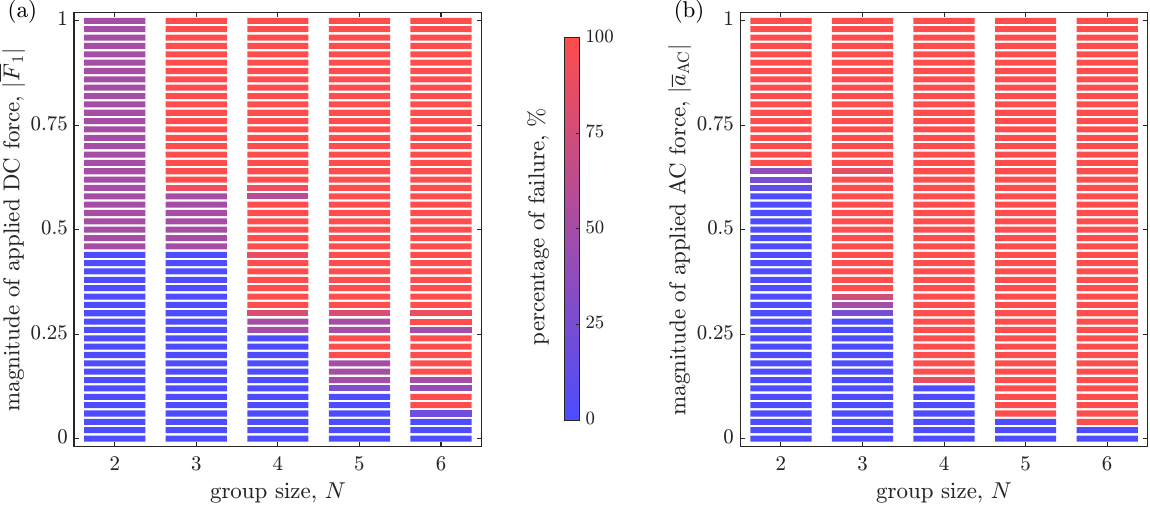}
    \caption{Percentage of group structure failure by collision or separation, for different external forcings and group sizes~$N$. (a) Impulse perturbation applied on the leader, as described by Eq.~\eqref{eq:ImpulseForceGaussian}. (b) Oscillatory forcing applied on the leader, as described by Eq.~\eqref{eq:acperturbation}. In both cases, we fix $\mathrm{Re}_f=1500$, $A^*=0.375$, and $M^*=0.16$, which correspond to the baseline case used in experiments~\cite{newbolt2024flow}. To assess the magnitude of the forces, we compare them against the self-thrust term $F_0=\rho C_Tcs(\pi Af)^2/2$, which after non-dimensionalization becomes $\pi^2C_TA^*/2$. Thus, in (a) and (b), the vertical axes are $\overline{F}_1=F_1/F_0$ and $\overline{a}_{\mathrm{AC}}=a_{\mathrm{AC}}/F_0$, respectively.
}
    \label{fig:disturbanceVsNmax}
\end{figure}

When external stresses, such as compression or stretching, are applied to a molecular crystal they can cause it to deform in a way that reflects the collective response of its atoms. This can potentially lead to plastic deformation or even fracture of the crystal. Similarly, external perturbations are expected to deform flying formations and, if unchecked, may cause collisions or break up the group. We model this scenario by applying two types of external perturbations to the leader in a group: an impulse and an oscillatory force. In doing so, we compute the percentage of failure for different group sizes. A failure here refers to group structure fracture either by collision or separation. In Figs.~\ref{fig:disturbanceVsNmax}(a) and \ref{fig:disturbanceVsNmax}(b), we fix the three dimensionless groups $M^*$, $\mathrm{Re}_f$, and $A^*$ as 0.16, 1500, and 0.375, respectively. The flyers are initially placed so that the dimensionless spacing between them is $S^*\approx 1+1/6$.

We first apply a sudden, brief force on the group leader. This takes the form of the impulse force perturbation described in Eq.~\eqref{eq:ImpulseForceGaussian}, and it corresponds to ``kicking'' the leader at $t\approx t_0$, where $t_0$ is chosen such that the flyers are already traveling at equilibrium. The time $t_0$ is therefore dependent on the mass, because the time it takes the flyers to reach equilibrium varies. 
We control the following quantities: The duration $\delta$ over which the kick acts, its magnitude $F_1$, and the direction of the kick. For each group size $N$ and force magnitude $F_1$, we allow the kick on the leader to be directed both away and towards the second flyer, i.e.\ we extend Eq.~\eqref{eq:ImpulseForceGaussian} to allow for both $+$ and $-$ signs. We also vary the duration: $\delta\in \{2\Delta,2.5\Delta,\dots , 4\Delta\}$.
Since the time steps in \textsc{Matlab}'s \texttt{ddesd} solver are non-uniform, the duration $\delta$ of the impulse force must be carefully selected to ensure it is sufficiently long; otherwise it lies entirely between two consecutive times $t_n$ and $t_{n+1}$. We choose $\delta\geq 2\Delta$ with $\Delta =\max_n \Delta t_n$. 

The effect of an impulse on a flyer is a sudden change in its velocity and a momentary perturbation of its trajectory. In some cases, this can have detrimental effects on the cohesion of the flying formation. We show the percentage of failure in the space of $N$-$|\overline{F}_{1}|$ in Fig.~\ref{fig:disturbanceVsNmax}(a),
where $\overline{F}_1=F_1/F_0$ is used to assess the magnitude of the external forces $F_1$ in comparison to the self-thrust term $F_0=\pi^2C_TA^*/2$. The percentage of failure ranges from 0\% for a group that always maintains its crystalline structure to 100\% for a group whose members are guaranteed to collide or separate from their neighbors. This percentage is computed by taking the average of the number of collisions that occur over the 10 cases of kick-duration $\delta$ and kick-direction $+/-$. 
The upper right corner of Fig.~\ref{fig:disturbanceVsNmax}(a) indicates that 100\% of the imposed impulse perturbations lead to failure of the group. However, the flyers in the lower left corner of the $N$-$|\overline{F}_1|$ space can also recover and stabilize, returning to an equilibrium state while maintaining group cohesion. When $N=2$ (two flyer-system; first column of Fig.~\ref{fig:disturbanceVsNmax}(a)) and $|\overline{F}_1|\gtrsim 0.5$, about 50\% of the perturbations lead to separation of the leader from the follower and thus failure of the crystalline structure. The remaining half move together cohesively, settling back to a steady flight.

The second kind of perturbation is an oscillatory force applied on the leader, while it is moving at equilibrium, as in Eq.~\eqref{eq:acperturbation}. This external force continuously causes oscillations or repeated deviations in the flyer’s motion, across all frequencies.
Figure~\ref{fig:disturbanceVsNmax}(b) shows the percentage of group failure in $N$-$|\overline{a}_{\mathrm{AC}}|$ space, where $|\overline{a}_{\mathrm{AC}}|$ is the magnitude of the applied force relative to the self-thrust.  Here, we seek to test the worst-case scenario near resonance, and so we set the frequency $f_{\mathrm{AC}}$ of the imposed oscillatory perturbation equal to the dimensionless resonant frequency in Eq.~\eqref{eq:fRresonance}, matching the flyer's natural frequency of oscillation from its flapping motion. 
Given a combination of $N$ and $|\overline{a}_{\mathrm{AC}}|$, the only free variable is the phase difference $\phi$ between the flyer's flapping motion and the imposed perturbation. We use eight values for $\phi$ (i.e.\ $\phi\in\{0,\pi/4,\dots,7\pi/4\}$) and calculate the percentage of group failure by averaging the number of collisions across these eight phase lag values for each $N$ and $|\overline{a}_{\mathrm{AC}}|$ combination. The bottom right corner of Fig.~\ref{fig:disturbanceVsNmax}(b) shows that with $N=6$, the group is very fragile and even small amplitudes of the oscillatory perturbation can lead to failure. A group of $N=2$ flyers is instead significantly safer, such that  sufficiently large forces need to be applied to fracture the crystalline group structure. In terms of the analogy with mechanical systems, this corresponds to the material having a compressive and tensile strength beyond which the inter-flyer bond breaks.

In general, the AC perturbation appears to be more ``dangerous'' for the flyers compared to the impulse force, since the computed percentages of failure are higher in Fig.~\ref{fig:disturbanceVsNmax}(b) than in Fig.~\ref{fig:disturbanceVsNmax}(a), particularly when $N$ is large.
In summary, the crystallinity of the group becomes increasingly more fragile in response to stronger perturbations.

 \section{Conclusions and Discussion}

Our investigations promote a view of flapping flight formations and flocks as aerodynamically-mediated materials whose unique physical properties arise from high-Reynolds-number flow interactions. The modeling, computational, and analytical studies presented here focus on the structural and dynamical consequences of wake interactions in the idealized context of oscillating propulsors arranged in tandem, linear or columnar arrays. Our model flocks can be summarized as behaving like self-propelling crystals whose members tend to order into lattice positions but whose unique bonds lead to longitudinal waves that make the group dynamics highly sensitive to perturbations. We take these findings as showing the promise and merits of applying the framework of active matter physics to the collective locomotion of flying or swimming animals.  

Regarding spatial structure of the flock, the material view provides the useful analogy to atoms or molecules in a crystal. Our analysis of the model shows that lattices represent equilibrium configurations for ensembles undergoing tandem or in-line flight, and our simulations reproduce this outcome for small and moderate system sizes. The associated lattice constant is set by the periodic or wave-like wake flow, which determines special positions where thrust and drag come into balance. These findings are broadly consistent with Lighthill's conjecture that flow interactions in collective locomotion may induce ordered arrangements \cite{lighthill1975mathematical}. They also connect to Vicsek's model of flocking as an emergent phase of matter and Weihs' hydrodynamic model of fish schools as lattices \cite{vicsek1995novel,weihs1973hydromechanics,weihs1975some}. Within our model, the origin of the crystallinity lies in the inter-flyer bond, which is investigated here in detail for the two-flyer problem. The thrust on a follower is modified due to its interaction with the wake of an upstream neighbor, and the net effect is that of a mechanical spring whose rest length sets the equilibrium position and whose stiffness dictates the restoring action and stability \cite{ramananarivo2016flow,newbolt2024flow}. The spring-like bond between members provides elasticity to the group that allows it to deform under loads and accommodate compression and tension. By imposing force perturbations, we show that the system tolerates strains up to about 20\%. Beyond that, the system catastrophically loses its elastic response much like the failure of a yield-stress material that deforms plastically and which here causes abrupt loss of structural integrity and collisions among neighbors.

Cohesion together at well separated positions allows the group to move en masse, which is presumably the primary goal of flocking. While group locomotion is a collective outcome of our model, the flight speed itself is in all cases determined solely by the leader. The leader is in fact unaffected by the followers since the interactions are one way or diodic as dictated by the wake flows left by one member and encountered by the next. Interestingly, followers can lock in to the leader's pace even when the flapping kinematics differ across the members. For example, a slow follower can be pulled along behind a faster-flapping leader by taking up a particular location in the wake where the flow energy can be harvested \cite{liao2007review,heydari2021school,fang2025flowinteractionsforwardflight}. This gives the material a certain tolerance for individuality or variations among its members, a property that seems beneficial in the context of the biological systems that motivate this work.

The motions internally within the line formation are nuanced, collective phenomena. The dominant features of these dynamics are a new type of traveling longitudinal waves which are analogous to phonons as vibrational modes of a lattice and which we call ``flonons'' to emphasize their conveyance by flows. Their key property is the growth in oscillation amplitude for later members, which leads to large excursions from the lattice positions and which destabilizes the formation by causing collisions between individuals. Our perturbation studies show that flonons place fundamental limits on the size of formations that can survive without feedback control, with maximal groups consisting of roughly 4 to 9 flyers depending on conditions. The root cause of the self-amplification is the diodic or nonreciprocal nature of the bonds, which cause gains in vibrational energy that accumulate downstream. The resulting disruptions and group fragility suggest limitations to Lighthill's idea of passive stability from flow interactions, and similarly lost are the crystalline phase and lattice arrangement that are appealing within Vicsek's and Weihs' frameworks. These findings suggest the biological interpretation that birds must use active sensing and feedback response behaviors to achieve the long columnar formations seen in nature. Further, the time scales of the instability growth are measured here to be variously 2 to 20 wing strokes, and we postulate that this constrains the active response to be similarly fast to suppress the instability.

Flonons and their consequences could also be interpreted as being advantageous to the group. Inherent fragility could be regarded positively as sensitively responsive. Dynamical instability is often viewed as beneficial in locomotion systems for which agility or maneuverability is important. In this case, an intrinsically unstable formation may be able to quickly change flight speed or rapidly fracture into multiple subgroups, abilities that seem favorable in contexts such as response to a predator. Related, flonons represent a form of mechanical amplification that seem uniquely well suited to detecting perturbations. Even an infinitesimally small disturbance would be amplified to sensible levels for downstream members in the group. This suggests biological interpretations related to sensing predators, prey, flow disturbances, or other environmental perturbations to which collective reactions are important. Similarly, flonons provide a way for members to transmit and communicate information to followers. Any change in flight state is felt sometime later by others, and here we have characterized the speed of information transmission as reflecting both the advection effect associated with the flight speed and the wave effect associated with the sound speed \cite{chaikin1995principles,beatus2006phonons}.

The material perspective promoted here suggests many directions for future research. A natural extension of our model would consider the continuum limit appropriate to large flocks and where the members would be tracked as a density field that depends on space and time \cite{carrillo2014derivation,bruna2025lane}. In such case, we expect a partial differential equation describing the dynamics and whose properties could be connected to the discrete model presented here. Work along this direction could include numerical solutions, equilibrium and stability analysis, wave solutions and their dispersion relation, etc. Another direction would involve considering large flocks with variability across its members, for which connections with disordered materials could be exploited. Preliminary work here and elsewhere suggests that differences in flapping kinematics lead to loss of crystalline order but also disruption of flonon amplification \cite{newbolt2019flow, newbolt2024flow}. Future work could aim to more systematically assess randomized flocks towards the goal of suppressing the instability and maximizing the size and lifetime of groups. Similarly, defects in the crystal such as vacancies could be characterized for their effect on flonon transmission. Finally, one could envision combining the model of the physical interactions developed here with behavioral or social models for active sensing and feedback response, thereby producing flocks that act as ``smart'' materials. A primary goal of such work would be to suppress flonon instabilities, a result that could help to explain how birds achieve long and long-lived columnar formations.

\section*{Acknowledgments}
\noindent C.M. acknowledges funding support provided by a Courant Instructorship and Joseph B. Keller Postdoctoral Fellowship at the Courant Institute at NYU. This work was supported by the U.S. National Science Foundation through the grant DMS-1847955 to L.R.

\bibliographystyle{unsrt}  
\bibliography{biblio.bib} 

\end{document}